\documentclass[natbib]{svjour3}                     
\smartqed  
\usepackage{graphicx}
\usepackage{aps-bibstyle}  
\graphicspath{{./fig/}{./png/}}

\graphicspath{{./fig/}{./png/}}
\def\tautd{\tau_{\rm td}}
\def\tauto{\tau_{\rm to}}

\newcommand{\EQ}{\begin{equation}}
\newcommand{\EN}{\end{equation}}
\newcommand{\EQA}{\begin{eqnarray}}
\newcommand{\ENA}{\end{eqnarray}}

\newcommand{\Eq}[1]{Equation~(\ref{#1})}

\newcommand{\Sec}[1]{Section~\ref{#1}}

\newcommand{\Fig}[1]{Figure~\ref{#1}}

\newcommand{\bra}[1]{\langle #1\rangle}

{}
{}
{}
\newcommand{\meanemf}{\overline{\cal E} {}}

{}
{}
\newcommand{\meanEMF}{\overline{\mbox{\boldmath ${\cal E}$}}{}}{}
{}
{}
{}
{}
{}
\newcommand{\meanAA}{\overline{\mbox{\boldmath $A$}}{}}{}
\newcommand{\meanBB}{\overline{\mbox{\boldmath $B$}}{}}{}
{}
{}
\newcommand{\meanFFf}{\overline{\mbox{\boldmath $F$}}_{\rm f}{}}{}
{}
{}
{}
{}
{}
\newcommand{\meanJJ}{\overline{\mbox{\boldmath $J$}}{}}{}
{}
\newcommand{\meanUU}{\overline{\mbox{\boldmath $U$}}{}}{}

{}
{}
{}

\newcommand{\meanB}{\overline{B}}

\newcommand{\meanhf}{\overline{h}_{\rm f}}

\newcommand{\meanU}{\overline{U}}

\newcommand{\meanJ}{\overline{J}}

\newcommand{\hatBB}{\hat{\vec{B}}}
{}

{}
{}

\newcommand{\hatB}{\hat{B}}

\newcommand{\rr}{\mbox{\boldmath $r$} {}}
\newcommand{\xx}{\mbox{\boldmath $x$} {}}
\newcommand{\yy}{\mbox{\boldmath $y$} {}}

\newcommand{\vv}{\mbox{\boldmath $v$} {}}

\newcommand{\uu}{\mbox{\boldmath $u$} {}}
\newcommand{\UU}{\mbox{\boldmath $U$} {}}

\newcommand{\bb}{\mbox{\boldmath $b$} {}}
\newcommand{\BB}{\mbox{\boldmath $B$} {}}

\newcommand{\jj}{\mbox{\boldmath $j$} {}}

\newcommand{\AAA}{\mbox{\boldmath $A$} {}}
\newcommand{\aaaa}{\mbox{\boldmath $a$} {}}

\newcommand{\nab}{\mbox{\boldmath $\nabla$} {}}

\newcommand{\oo}{\mbox{\boldmath $\omega$} {}}

\newcommand{\ii}{{\rm i}}

\newcommand{\dd}{{\rm d} {}}

\def\Pm{\mbox{\rm Pr}_M}
\def\Rm{\mbox{\rm Re}_M}

\def\Rmc{R_{\rm m,{\rm crit}}}
\def\Rey{\mbox{\rm Re}}

\def\alpK{\alpha_{\rm K}}
\def\alpM{\alpha_{\rm M}}
\def\EK{E_{\rm K}}
\def\EM{E_{\rm M}}

\def\kmean{k_{\rm m}}

\def\kf{k_{\rm f}}
\def\HM{H_{\rm M}}
\def\EM{E_{\rm M}}
\def\epsf{\epsilon_{\rm f}}

\def\urms{u_{\rm rms}}

\def\etat{\eta_{\rm t}}
\def\etatz{\eta_{\rm t0}}

\def\Beq{B_{\rm eq}}
\def\Beqz{B_{\rm eq0}}
\def\tautd{\tau_{\rm td}}
\def\tauto{\tau_{\rm to}}

\def\half{{\textstyle{1\over2}}}

\def\onethird{{\textstyle{1\over3}}}

\newcommand{\yan}[5]{, #5, {Astron.\ Nachr.\ }{\bf #2}, #3-#4 (#1)}

\newcommand{\yana}[5]{, #5, {Astron.\ Astrophys.\ }{\bf #2}, #3-#4 (#1)}
\newcommand{\yanaN}[4]{, #4, {Astron.\ Astrophys.\ }{\bf #2}, #3 (#1)}

\newcommand{\yass}[5]{, #5, {Astrophys.\ Spa.\ Sci.\ }{\bf #2}, #3-#4 (#1)}

\newcommand{\psph}[3]{, #2, {Solar Phys.}, #3 (#1)}
\newcommand{\ysph}[5]{, #5, {Solar Phys.\ }{\bf #2}, #3-#4 (#1)}
\newcommand{\yjetp}[5]{, #5, {Sov.\ Phys.\ JETP }{\bf #2}, #3-#4 (#1)}
\newcommand{\yspd}[5]{, #5, {Sov.\ Phys.\ Dokl.\ }{\bf #2}, #3-#4 (#1)}

\newcommand{\ysovl}[5]{, #5, {Sov.\ Astron.\ Letters }{\bf #2}, #3-#4 (#1)}
\newcommand{\ymn}[5]{, #5, {Monthly Notices Roy.\ Astron.\ Soc.\ }{\bf #2}, #3-#4 (#1)}
\newcommand{\pmn}[3]{, #2, {Monthly Notices Roy.\ Astron.\ Soc.\ }#3 (#1)}

\newcommand{\ynat}[5]{, #5, {Nature }{\bf #2}, #3-#4 (#1)}

\newcommand{\yjfm}[5]{, #5, {J.\ Fluid Mech.\ }{\bf #2}, #3-#4 (#1)}
\newcommand{\ypr}[5]{, #5, {Phys.\ Rev.\ }{\bf #2}, #3-#4 (#1)}
\newcommand{\ypre}[5]{, #5, {Phys.\ Rev.\ E }{\bf #2}, #3-#4 (#1)}
\newcommand{\ypreN}[4]{, #4, {Phys.\ Rev.\ }{\bf #2}, #3 (#1)}
\newcommand{\yprl}[5]{, #5, {Phys.\ Rev.\ Letters }{\bf #2}, #3-#4 (#1)}
\newcommand{\yprlN}[4]{, #4, {Phys.\ Rev.\ Letters }{\bf #2}, #3 (#1)}

\newcommand{\yprs}[5]{, #5, {Proc.\ Roy.\ Soc.\ Lond.\ }{\bf #2}, #3-#4 (#1)}
\newcommand{\yptrs}[5]{, #5, {Phil.\ Trans.\ Roy.\ Soc.\ }{\bf #2}, #3-#4 (#1)}

\newcommand{\yapj}[5]{, #5, {Astrophys.\ J.\ }{\bf #2}, #3-#4 (#1)}
\newcommand{\papj}[3]{, #2, {Astrophys.\ J.}, arXiv:#3 (#1)}
\newcommand{\yapjN}[4]{, #4, {Astrophys.\ J.\ }{\bf #2}, #3 (#1)}

\newcommand{\yapjl}[5]{, #5, {Astrophys.\ J.\ Letters }{\bf #2}, #3-#4 (#1)}

\newcommand{\yppN}[4]{, #4, {Phys.\ Plasmas }{\bf #2}, #3 (#1)}

\newcommand{\yaraa}[5]{, #5, {Ann.\ Rev.\ Astron.\ Astrophys.\ }{\bf #2}, #3-#4 (#1)}
\newcommand{\yrmp}[5]{, #5, {Rev.\ Mod.\ Phys.\ }{\bf #2}, #3-#4 (#1)}

\newcommand{\yanf}[5]{, #5, {Ann.\ Rev.\ Fluid Dyn.\ }{\bf #2}, #3-#4 (#1)}
\newcommand{\ypf}[5]{, #5, {Phys.\ Fluids }{\bf #2}, #3-#4 (#1)}
\newcommand{\ypfN}[4]{, #4, {Phys.\ Fluids }{\bf #2}, #3 (#1)}
\newcommand{\yphy}[5]{, #5, {Physica } {\bf #2}, #3-#4 (#1)}
\newcommand{\ygafd}[5]{, #5, {Geophys.\ Astrophys.\ Fluid Dynam. }{\bf #2}, #3-#4 (#1)}

\newcommand{\yrppN}[4]{, #4, {Rep.\ Progr.\ Phys.\ }{\bf #2}, #3 (#1)}

\newcommand{\ypfb}[5]{, #5, {Phys.\ Fluids B }{\bf #2}, #3-#4 (#1)}
\newcommand{\yjour}[6]{, #6, {#2} {\bf #3}, #4-#5 (#1)}
\newcommand{\yjourN}[5]{, #5, {#2} {\bf #3}, #4 (#1)}

\newcommand{\ybook}[3]{ {#2}.\ #3 (#1)}

\begin{document}

\title{Current status of turbulent dynamo theory}
\subtitle{From large-scale to small-scale dynamos}
\titlerunning{Current status of turbulent dynamo theory}

\author{Axel Brandenburg, Dmitry Sokoloff, Kandaswamy Subramanian}
\authorrunning{Brandenburg et al.}

\institute{A. Brandenburg \at
Nordita, Royal Institute of Technology and Stockholm University,
Roslagstullsbacken 23, 10691 Stockholm, Sweden; and
Department of Astronomy, Stockholm University, SE 10691 Stockholm, Sweden,
\email{brandenb@nordita.org}
\and
D. Sokoloff \at
Department of Physics, Moscow University, 119992 Moscow, Russia,
\email{d\_sokoloff@hotmail.com}
\and
K. Subramanian \at
Inter-University Centre for Astronomy and Astrophysics,
Post Bag 4, Ganeshkhind, Pune 411 007, India,
\email{kandu@iucaa.ernet.in}
}

\date{\today,~ $ $Revision: 1.138 $ $}

\maketitle

\begin{abstract}
Several recent advances in turbulent dynamo theory are reviewed.
High resolution simulations of small-scale and large-scale dynamo
action in periodic domains are compared with each other and contrasted
with similar results at low magnetic Prandtl numbers.
It is argued that all the different cases show similarities at
intermediate length scales.
On the other hand, in the presence of helicity of the turbulence,
power develops on large scales, which is not present in non-helical 
small-scale turbulent dynamos.
At small length scales, differences occur in connection with the
dissipation cutoff scales associated with the respective value
of the magnetic Prandtl number.
These differences are found to be independent of whether or not there
is large-scale dynamo action.
However, large-scale dynamos in homogeneous systems are shown to suffer
from resistive slow-down even at intermediate length scales.
The results from simulations are connected to mean field theory and
its applications. Recent work on helicity fluxes to alleviate
large-scale dynamo quenching, shear dynamos, nonlocal effects and
magnetic structures from strong density stratification are highlighted.
Several insights which arise from analytic considerations 
of small-scale dynamos are discussed.

\end{abstract}

\section{Introduction}
\label{intro}

Dynamos convert kinetic energy into magnetic energy.
In the astrophysical context, one always means by a
dynamo a self-excited device, except that
the conductivity is uniformly
distributed and not limited to conducting wires in an
insulating container, as in laboratory dynamos.
Since the beginning of the space age, it is well understood
that such a device can work, at least in principle.
The discovery of the first rigorously proven example by \cite{Her58}
was significant, even though the particular configuration studied by
him was not of immediate astrophysical interest.
The Herzenberg configuration consists of a uniformly conducting
solid medium (e.g., copper) with two or more conducting rotors
in electrical contact with the rest.
Dynamo action is possible when the rotors spin faster than a certain
critical value that depends on the angle between the rotors.
For angles between 90 and 180 degrees the solutions are non-oscillatory,
while for angles between 0 and 90 degrees there are oscillatory solutions
that were discovered only more recently \citep{BMS98}.

The Herzenberg dynamo belongs to the class of ``slow'' dynamos, for
which the dynamo growth rate is maximum for intermediate values of the
conductivity, but it returns to zero in the limit of perfect conductivity.
Another example of a slow dynamo is the Roberts flow, which consists of
a steady two-dimensional flow pattern, $\uu=\uu(x,y)$, but all three
flow components are non-vanishing.
This dynamo was first studied by \cite{Rob70,Rob72}.
It provides an important benchmark of a large-scale dynamo,
where magnetic field on scales larger than the scale of motions are produced,
by a mechanism called the $\alpha$ effect.
Here, $\alpha$ refers to the name of the coefficient in the relation between
mean magnetic field $\meanBB$ and mean electromotive force $\meanEMF$
in a turbulent flow.
The basic idea goes back to \cite{Par55} who proposed a relation of the
form $\meanEMF=\alpha\meanBB$.
Later, \cite{SKR66} computed a tensorial relation of the form
$\meanemf_i=\alpha_{ij}\meanB_j$ for rotating stratified turbulence.
It was clear that higher derivatives of the magnetic field would also enter,
so a more general expression is
\EQ
\meanemf_i=\alpha_{ij}\circ\meanB_j+\eta_{ijk}\circ\meanB_{j,k},
\label{meanemf}
\EN
where the isotropic part of the tensor $\eta_{ijk}$,
$\eta_{ijk}=\etat\epsilon_{ijk}$, corresponds to turbulent magnetic,
diffusion where $\etat$ is the {\it turbulent} magnetic diffusivity.
The circles indicate convolution over time and space,
which is however commonly replaced by a multiplication
in the limit in which the integral kernels are well approximated
by $\delta$ functions in space and time.
The work of \cite{SKR66} marked the beginning of mean-field electrodynamics,
which is still the leading theory to explain the amplification of magnetic
fields on length scales that are larger than the scale of
the energy-carrying turbulent eddies.
Such systems are therefore also referred to as large-scale dynamos.
Essential here is that the trace of the $\alpha$ tensor is non-vanishing.
This typically requires helicity of streamlines (at low conductivity)
or of vortex lines (at high conductivity).

In this review we discuss our current knowledge of dynamos
covering both large-scale and small-scale turbulent dynamos.
In practice, all dynamos of astrophysical relevance tend to be ``fast''
and thus have a finite growth rate also in the limit of perfect conductivity.
There are examples of predetermined flows for which fast dynamo action is
indicated by numerical simulations, for example the ABC flow \citep{GF86}
and the \cite{GP92} flow, but the convergence of the $\alpha$
effect with increasing magnetic Reynolds number can be quite slow
or non-existent \citep[see, e.g.,][]{CHT06}.
By contrast, when relaxing the assumption of a predetermined kinematic
flow pattern and adopting a turbulent flow field, the $\alpha$ effect
appears to be converged for magnetic Reynolds numbers exceeding a critical
value of the order of unity
up to values of about 200 probed in the simulations \citep{SBS08}.

We begin by discussing examples of numerical realizations of
turbulent dynamos and then turn to some astrophysical examples.
Large-scale dynamos produce fields that are well characterized by
suitable averages.
A theory for describing the evolution of such averaged fields is
mean-field theory that is obtained by averaging the governing equations,
most notably the induction equation.
Large-scale dynamos can then also be referred to as mean-field dynamos (MFD).
Under certain conditions, making suitable assumptions about the spatial
variation of the $\alpha$ effect, solutions of the mean-field induction
equation have been used to characterize the large-scale
magnetic fields seen in astrophysical bodies like the Sun and some
spiral galaxies.
Mean-field theory is also applied to the momentum equation.
This leads to a number of effects including turbulent viscosity,
the $\Lambda$ effect \citep[responsible driving differential solar
rotation; see][]{Rue80,Rue89}, the anisotropic kinetic alpha effect
\citep{Frisch}, as well as the negative effective magnetic pressure
effect \citep{KRR89}.
In this paper, we also
discuss important effects resulting from the mean-field
momentum equation, namely the spontaneous formation of magnetic flux
concentrations in a strongly stratified layer.
Such results may be relevant to explaining the formation of active regions
in the Sun.

There is now increased interest in what is known as small-scale or
fluctuation dynamos.
These are dynamos that can work already under fully isotropic conditions
and were anticipated by \cite{Bat50} and others \citep{BS51,Els56},
but the now accepted theory was provided by \cite{Kaz68}.
Fluctuation (or small-scale) dynamos are important in cosmic objects
because they are generic to any random flow of a sufficiently
conducting plasma.
Furthermore, the growth rate is fast and random fields can grow
on the eddy turnover timescale of the smallest eddies, which are
typically much shorter than the age of the system.
This is particularly true of galaxy clusters,
for which small-scale dynamos are crucial to explaining the observed
magnetic fields, because the conditions for large-scale dynamo action are
probably absent \citep{SSH06}.

Small-scale dynamos are nowadays also invoked to describe the small-scale
magnetic field at the solar surface.
However, in many contexts both small-scale and large-scale dynamos go together.
And then it is not clear whether one can (or should) distinguish between
a small-scale field from a small-scale dynamo from that associated with
the fluctuations that are inherent to any large-scale dynamo and that
can be caused by tangling and amplification of the large-scale field.

\section{Numerical realizations of turbulent dynamos}
\label{NumericalRealizations}

Next, we discuss some numerical realizations of turbulent dynamos.
By ``turbulence'' we mean here flows that are solutions to the
Navier-Stokes equations that are irregular in space and time, subject
to energy injection at large length scales and energy dissipation at
short length scales.
The ratio between forcing scale to dissipation scale is quantified by the
Reynolds number.
Throughout this paper, we adopt the following definition for this number:
\EQ
\Rey=\urms/\nu\kf,
\EN
where $\urms=\bra{\uu^2}^{1/2}$ is the root mean square (rms) value
of the turbulent velocity, $\nu$ is the kinematic viscosity, and $\kf$
is the forcing wavenumber, i.e., the wavenumber of energy injection.
A similar definition is adopted for the {\it magnetic} Reynolds number,
$\Rm=\urms/\eta\kf$, where $\eta$ is the magnetic diffusivity.
We also define the magnetic Prandtl number, $\Pm=\nu/\eta=\Rm/\Rey$.

At small Reynolds numbers, $\Rey\ll1$, the flow is determined by viscosity
and has qualitatively different properties from flows with $\Rey\gg1$.
The break point is close to unity, so, for example, a flow with $\Rey=5$
would be called turbulent (or at least mildly turbulent), because it begins
to show certain asymptotic scaling properties that are also found for
fully turbulent flows.
The most famous aspect of fluid turbulence
is the Kolmogorov $\EK(k)\sim k^{-5/3}$
energy spectrum, which is normalized such that $\int\EK\,\dd k=\urms^2/2$.
Density is omitted in this definition, which requires that density
fluctuations are unimportant and the flow is nearly incompressible.
In this review we also discuss stratified flows in which density
varies strongly due to gravity.
In those cases the incompressibility condition $\nab\cdot\uu=0$ has
to be replaced by $\nab\cdot\rho\uu=0$, which is valid as long
as the flows are slow compared with the sound speed and the typical
scales of variation smaller than a scale height.
However, in the following we do not make such assumptions and
consider fully compressible flows.

\subsection{Small-scale and large-scale dynamos}

\begin{figure}[t!]\begin{center}
\includegraphics[width=\textwidth]{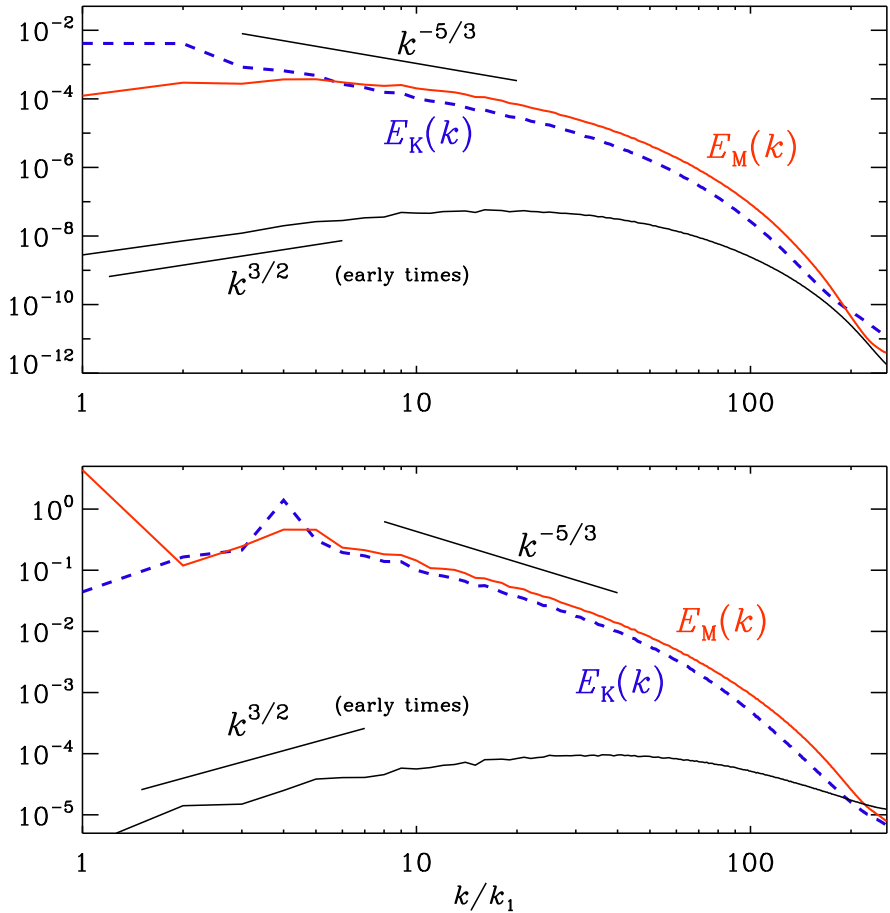}
\end{center}\caption[]{
Kinetic and magnetic energy spectra in a turbulence simulation without
net helicity (up) and with net helicity (bottom) for a magnetic Prandtl
number of unity and a mesh size is $512^3$ meshpoints.
Notice the pronounced peak of $\EM(k)$ at $k=k_1$ in the case with helicity.
The energy input wavenumbers are $\kf=1.5 k_1$ in the non-helical case
(upper panel, $\Rm=600$, $\Pm=1$)
and $\kf/k_1=4$ in the helical case (lower plot, $\Rm=450$, $\Pm=1$).
Adapted from \cite{BS05} and \cite{BRRS08}.
}\label{pspec_nohel512d2_pspec_PrM1}\end{figure}

We begin by discussing first the difference between small-scale
and large-scale dynamos.
A small-scale dynamo is one that generates magnetic field
at scales much smaller than that of the energy-carrying eddies,
while a large-scale dynamo generates field at scales larger than
that of the energy-carrying eddies.
In may practical situations, the difference is of somewhat academic
interest, because the small-scale dynamo is always excited when the
magnetic Reynolds number is large, which is the case in many astrophysical
situations.
Furthermore, the conditions for the excitation of large-scale dynamo
action are met in many situations of interest.
However, in the case of isotropic turbulence such a distinction
can be made by considering helical and non-helical turbulence.
In both cases the system can be homogeneous and it makes sense to
compute spectra of magnetic and kinetic energy, $\EM(k)$ and $\EK(k)$,
respectively; see \Fig{pspec_nohel512d2_pspec_PrM1} for $\Pm=1$.
The early evolution of such a dynamo is quite similar:
both dynamos have a $k^{3/2}$ power spectrum at small scales.
Such a scaling was predicted by \cite{Kaz68} (see also \cite{KA92}) 
for a single scale non-helical 
flow which was $\delta$-correlated in time, but seems to be obtained
in the simulations more generally.
At late times, however, helical turbulence allows the development
of an inverse cascade 
\citep{Frisch75} on a longer resistive timescale (see below).

We see that, at least in the saturated state, large-scale dynamos
produce and sustain magnetic fields at scales larger than the energy
injection scale, while small-scale dynamos produce and sustain magnetic
fields at scales smaller than the energy injection scale.
The lack of similar behavior in the linear regime could be interpreted
as evidence that the underlying mechanism for producing large-scale fields
must be nonlinear in nature \citep{Rincon}.
Alternatively, one could interpret the resulting dynamo as a combination
of large-scale and small-scale dynamo action, where the latter has
a larger growth rate such that in the kinematic regime the field is
dominated by the small-scale field, although the large-scale dynamo
does still operate in the background.
As a consequence it gets the chance to dominate
only when the small-scale dynamo has already saturated.
Even in this case, and if the $\Rm$ is large,
we will argue that one needs to additionally
dissipate small-scale magnetic helicity before the largest scale
field can appear.

\begin{figure}[t!]\begin{center}
\includegraphics[width=\textwidth]{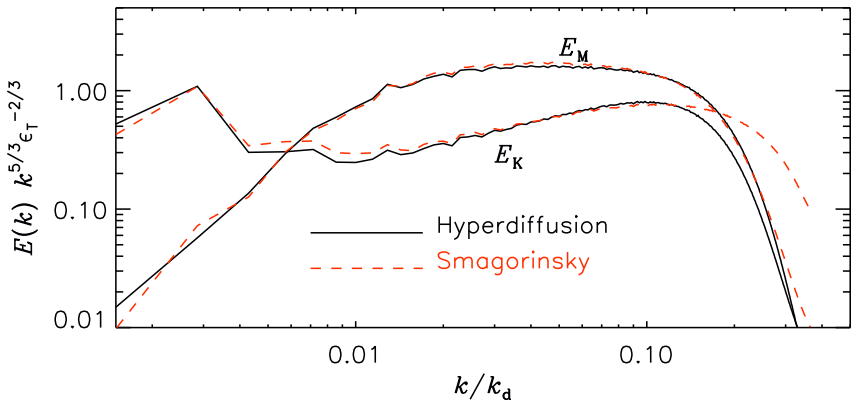}
\end{center}\caption[]{
Magnetic and kinetic energy spectra for runs with $512^3$
meshpoints and hyperviscosity with hyperresistivity (solid line) and
Smagorinsky viscosity with hyperresistivity (red, dashed line).
Note the mutual approach of kinetic and magnetic energy spectra
before entering the dissipative subrange.
Adapted from \cite{HB06}.
}\label{hyper_512}\end{figure}

Both large-scale and small-scale dynamos show that the spectral
magnetic energy exceeds the kinetic spectral energy in the beginning
of the inertial range.
This difference is slightly weaker for large-scale dynamos,
but this might be an artifact of the Reynolds number
still not being large enough.
Such a difference used to be completely absent at Reynolds numbers
previously reported; see, for example \cite{B01}.
The slight super-equipartition was quite evident when simulations at a
resolution of $1024^3$ meshpoints became first available \citep{HBD03},
although this feature can already be seen in earlier simulations
\citep{MFP81,KYM91}.
This spectral excess of magnetic fields is expected to diminish as one
proceeds further down the inertial range.
Such behavior was indeed seen in simulations of \cite{HB06} when using
a combination of Smagorinsky viscosity for the velocity field and
hyperresistivity for the magnetic field; see \Fig{hyper_512}.
This implies that the two spectra cannot be parallel to each other
at intermediate length scales, and that the slope of $\EM(k)$ must be
slightly steeper than that of $\EK(k)$.
This difference in the two slopes at intermediate wavenumbers is now
associated with the observed differences in the spectral exponents
in the solar wind; see \cite{Boldyrev}, who find steeper spectra for the
magnetic field than the velocity field both from simulations and solar
wind observations.

\begin{figure}[t!]\begin{center}
\includegraphics[width=\columnwidth]{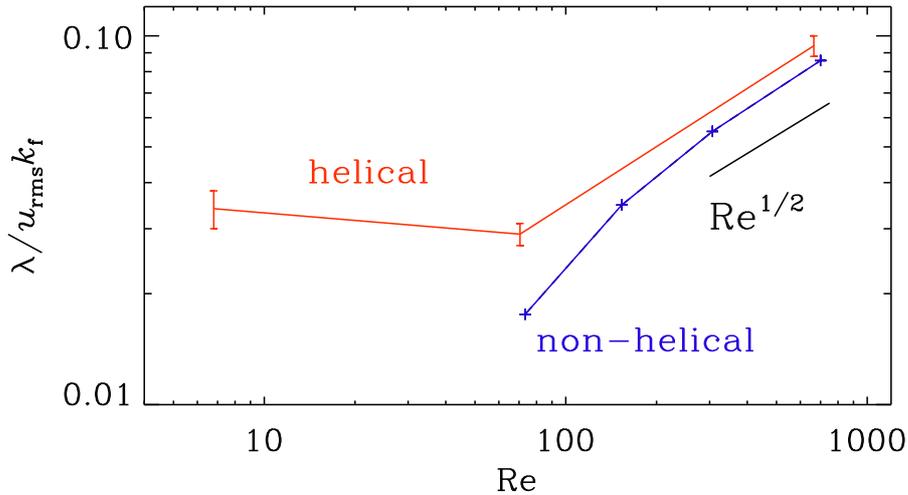}
\end{center}\caption[]{
Dependence of dynamo growth rates of the rms magnetic field
on $\Rm$ for helical and non-helical turbulence.
Adapted from \cite{B09}.
}\label{pgrowth}\end{figure}

In the spectra of \Fig{pspec_nohel512d2_pspec_PrM1} we 
see a remarkable difference between small-scale and large-scale dynamos.
In particular, large-scale dynamos are capable of producing a peak of
magnetic energy at the smallest possible wavenumber, while small-scale
dynamos do not.
On the other hand, we have stated earlier that small-scale and large-scale
dynamos are difficult to distinguish in the early stage.
In \Fig{pgrowth} we show the critical values of $\Rm$ ($=\Rey$) for dynamo
action both for small-scale (non-helical) and large-scale (helical) dynamos.
Note that for $\Rm>35$ the growth rate attains a $\Rey^{1/2}$ scaling
both for small-scale and large-scale dynamos.
The same growth rates are also obtained for dynamos driven by convection;
see Fig.~15 of \cite{KKB08}.
This $\Rey^{1/2}$ scaling of the growth rate of the rms magnetic field,
implies that the growth rate
is not controlled by the turnover time of the energy-carrying eddies,
but of eddies at the dissipation scale \citep{Scheko02,Scheko04}.
At low values of $\Rm$, only large-scale dynamo action remains possible.
Its excitation condition is determined by the requirement that a certain
dynamo number exceeds a critical value.
This usually translates to the condition that the degree of scale separation
is large enough; see equation~(5) of \cite{B09}.

An important issue for astrophysical applications is how coherent
are the fields generated by small-scale dynamos \citep{SSH06,CR09}. 
On the basis of simulations done with large $\Pm$, but small $\Rey$,
\cite{Schek04b} argued that the field saturates with a folded structure,
where the field reverses at the folds such that power concentrates on
resistive scales.
The simulations of \cite{HBD03,HBD04} with $\Rm=\Rey\gg1$,
found the magnetic correlation lengths $\sim 1/6$th the velocity
correlation length, but much larger than the resistive scale.
This seems consistent with the simple \cite{S99}
model for nonlinear saturation of small-scale dynamos.
What happens at large $\Rey$ and large $\Pm$, which is representative
of galactic and cluster plasmas, is not easy to capture in simulations.
The $\Pm=50$, $\Rey=80$ simulation described in \cite{BS05}, showed
evidence for strong field regions with folds, but equally, regions
with strong fields and no folds, illustrating that such structures
need not be volume filling. 
Moreover, the `spontaneous stochasticity', that applies to highly turbulent
flows \citep{Eyink}, suggests that the dynamics of small-scale dynamos could
be quite different in turbulent compared to laminar high-$\Pm$ systems. 
Furthermore, in galaxy clusters the viscosity may be set by
plasma effects \citep{Scheko05}. 
It could also be highly
anisotropic owing to the presence of magnetic fields \citep{Petal12}.
In addition, heat condition is also very anisotropic, giving rise
to magnetothermal and heat flux-driven buoyancy instabilities
\citep{PS05,PQ08}.
Further work on these aspects is desirable, using both semi-analytical
ideas and high resolution simulations.

\subsection{Low magnetic Prandtl number}
\label{Low}

\begin{figure}[t!]\begin{center}
\includegraphics[width=\columnwidth]{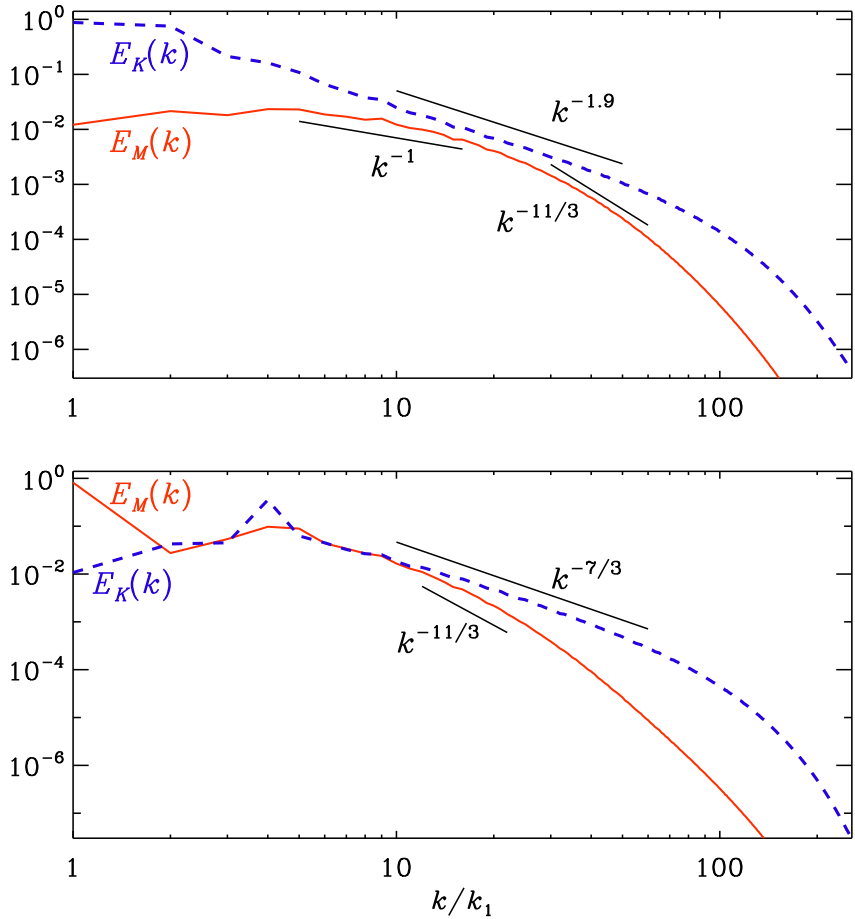}
\end{center}\caption[]{
Kinetic and magnetic energy spectra for non-helical (top) and
helical (bottom) turbulence at low magnetic Prandtl numbers of
$\Pm=0.02$ and 0.01, respectively.
Here, $\Rm=230$ with $\Rey=11,500$ in the non-helical case,
adapted from \cite{B11},
and $\Rm=23$ with $\Rey=2300$ in the helical case
adapted from \cite{B09}.
In both cases the resolution is $512^3$ mesh points.
}\label{pspec2}\end{figure}

In many astrophysical settings,
like solar, stellar, or accretion disk plasmas,
the magnetic Prandtl number is rather
low ($\sim10^{-4}$ or less).
This becomes numerically hard to handle, especially if the magnetic
Reynolds number should still be large enough to support dynamo action.
In \Fig{pspec2} we compare non-helical and helical runs with
magnetic Prandtl numbers of 0.02 and 0.01, respectively.
In the former case the magnetic Reynolds number
$\Rm$ is 230, which is
weakly supercritical; the critical value of $\Rm$
for small values of $\Pm$ is
$\Rmc\approx150$ \citep{B11}, compared with $\Rmc\approx35$ for $\Pm=1$
\citep{HBD04}.
These values agree with those of earlier work \citep{Scheko05b,Scheko07,Iska}.
So, for $\Rm=230$ and $\Pm=0.02$, we have $\Rey=\Rm/\Pm=11,500$.
Normally, this would require a numerical resolution of about $10,000^3$
meshpoints, but it turns out that at low values of $\Pm$, most of the
energy is dissipated resistively, leaving thus very little kinetic energy
to be cascaded, terminating therefore the kinetic energy cascade
earlier than at $\Pm=1$.

\begin{figure}[t!]\begin{center}
\includegraphics[width=.95\columnwidth]{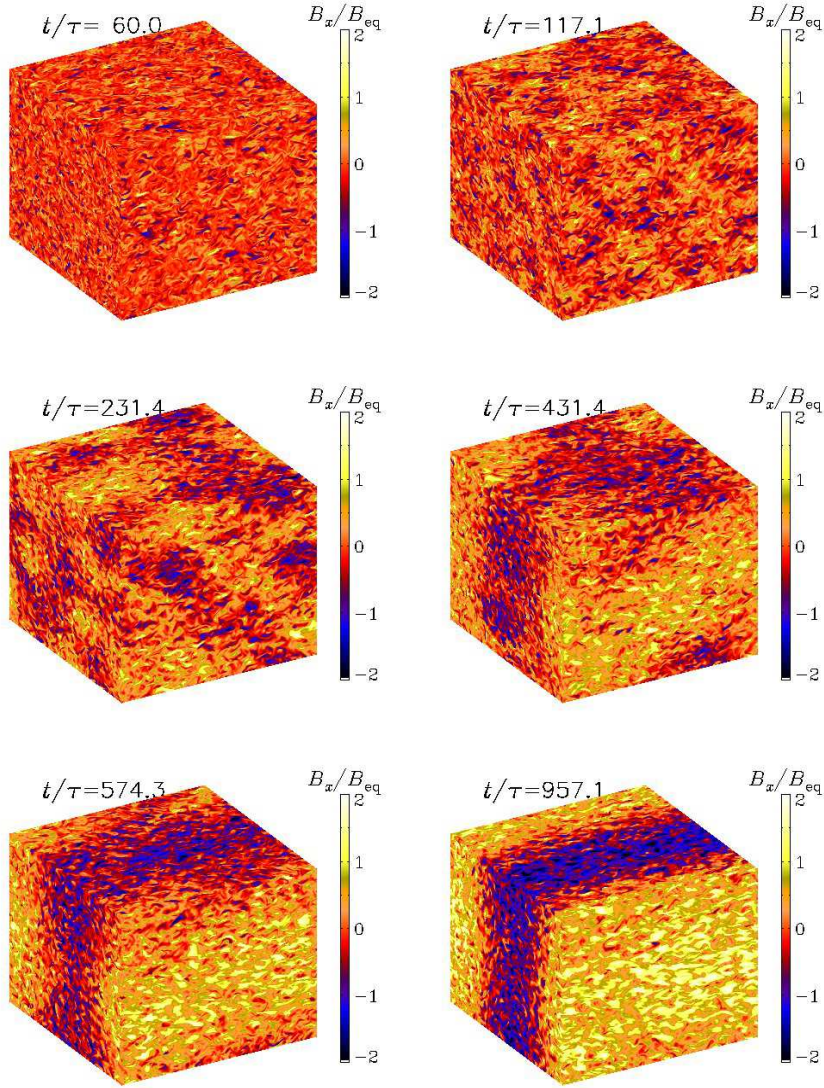}
\end{center}\caption[]{
Visualizations of $B_x/\Beq$ on the periphery of the domain at six
times during the late saturation stage of the dynamo when a large-scale
field is gradually building up.
The small-scale field has reached its final value after $t/\tau\approx100$
turnover times.
Here, $B_{\rm eq}=\sqrt{\mu_0\rho_0}\,\urms$ is the equipartition
field strength where kinetic and magnetic energy densities are comparable,
and $\rho_0$ is the mean density.
Note that the maximum field strength is about twice $B_{\rm eq}$.
}\label{B}\end{figure}

\begin{figure}[t!]\begin{center}
\includegraphics[width=\columnwidth]{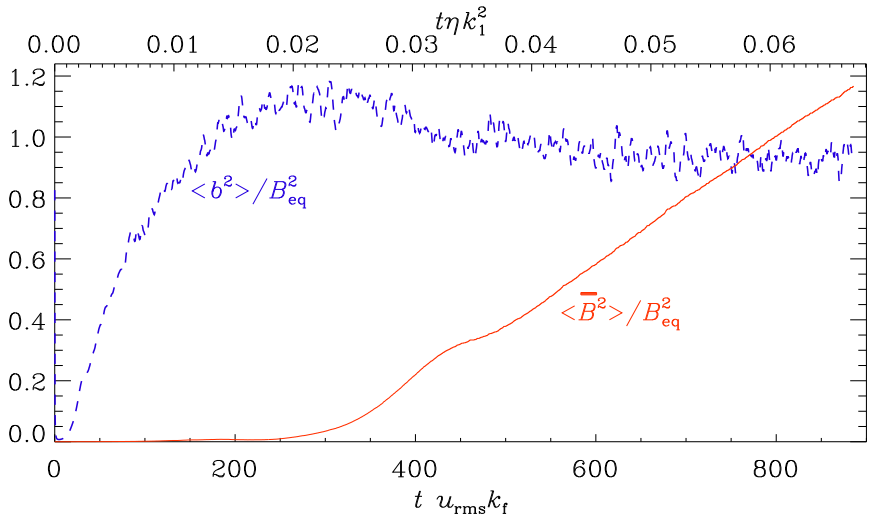}
\end{center}\caption[]{
Saturation of the small-scale magnetic energy density and continued
increase of the large-scale magnetic energy density.
Here, $\kf/k_1=15$.
}\label{psat}\end{figure}

At intermediate length scales, kinetic and magnetic energy spectra
are close to each other.
The magnetic energy spectrum no longer exceeds the kinetic energy
spectrum, as was found for $\Pm=1$; see \Fig{pspec_nohel512d2_pspec_PrM1}.
Again, this might be a consequence of still insufficiently large
Reynolds numbers and limited resolution.
In fact, it is plausible that, in the limit of large fluid and
magnetic Reynolds numbers, kinetic and magnetic energy spectra
coincide, even if $\Pm$ is small. And only at much smaller scales, 
the magnetic energy spectrum turns into a dissipative 
subrange, and goes below the kinetic energy spectrum, due
to stronger Ohmic dissipation.
The slopes of the $k^{-11/3}$ spectrum of \cite{Gol60} and
\cite{Mof61} and the scale-invariant $k^{-1}$ spectrum
\citep{RS82,KR94,KMR96} are shown for comparison.
However, the kinetic energy spectrum tends to be steeper than
$k^{-5/3}$ and is closer to $k^{-1.9}$ and $k^{-7/3}$ in the
non-helical and helical cases.

\begin{figure}[t!]\begin{center}
\includegraphics[width=\columnwidth]{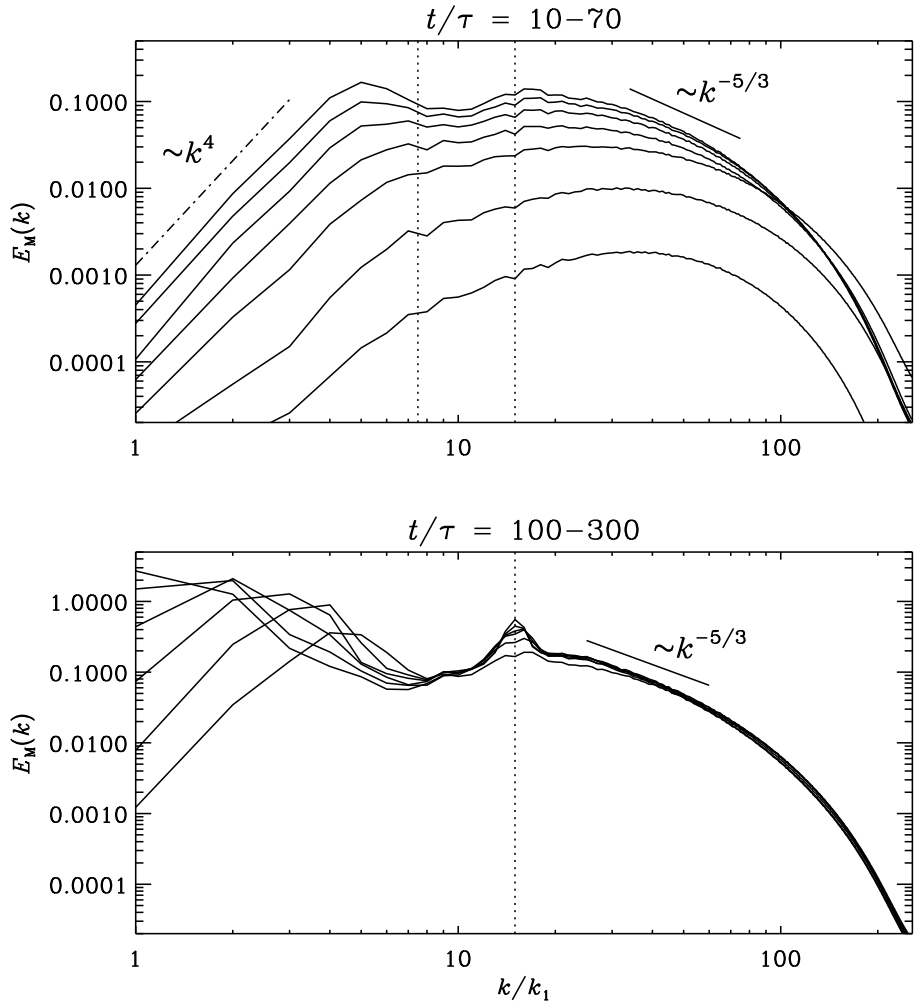}
\end{center}\caption[]{
Magnetic energy spectra $\EM(k)$, at earlier (top) and later (bottom) times.
The scale separation ratio is $\kf/k_1=15$.
The range of time $t$ is given in  units of the turnover time,
$\tau=1/\urms\kf$.
At small wavenumbers, the $\EM(k)$ spectrum is proportional to $k^4$,
while to the right of $\kf/k_1=15$ there is a short range with
a $k^{-5/3}$ spectrum.
}\label{pspec_ck}\end{figure}

\begin{figure}[t!]\begin{center}
\includegraphics[width=\columnwidth]{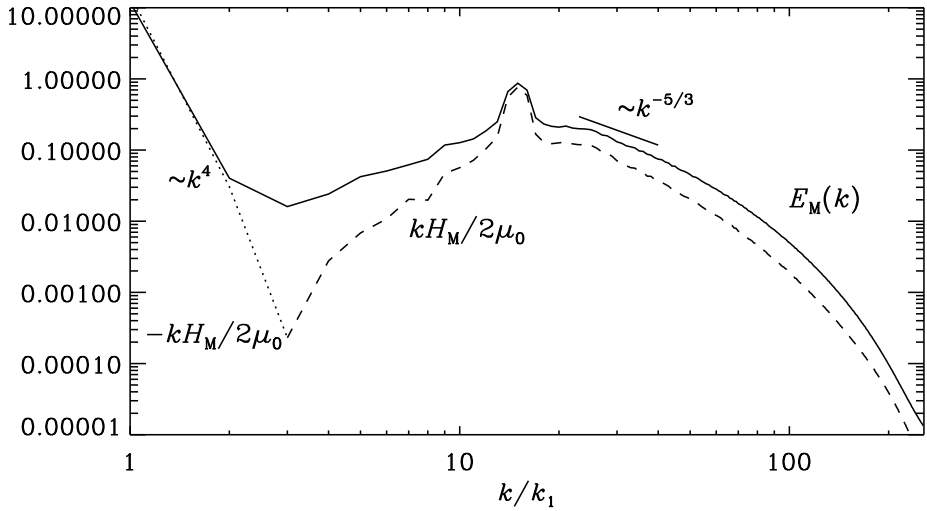}
\end{center}\caption[]{
Spectra of magnetic energy, $\EM(k)$,
and rescaled magnetic helicity, $\pm k\HM(k)/2\mu_0$.
}\label{pspec_ck0}\end{figure}

\begin{figure}[t!]\begin{center}
\includegraphics[width=\columnwidth]{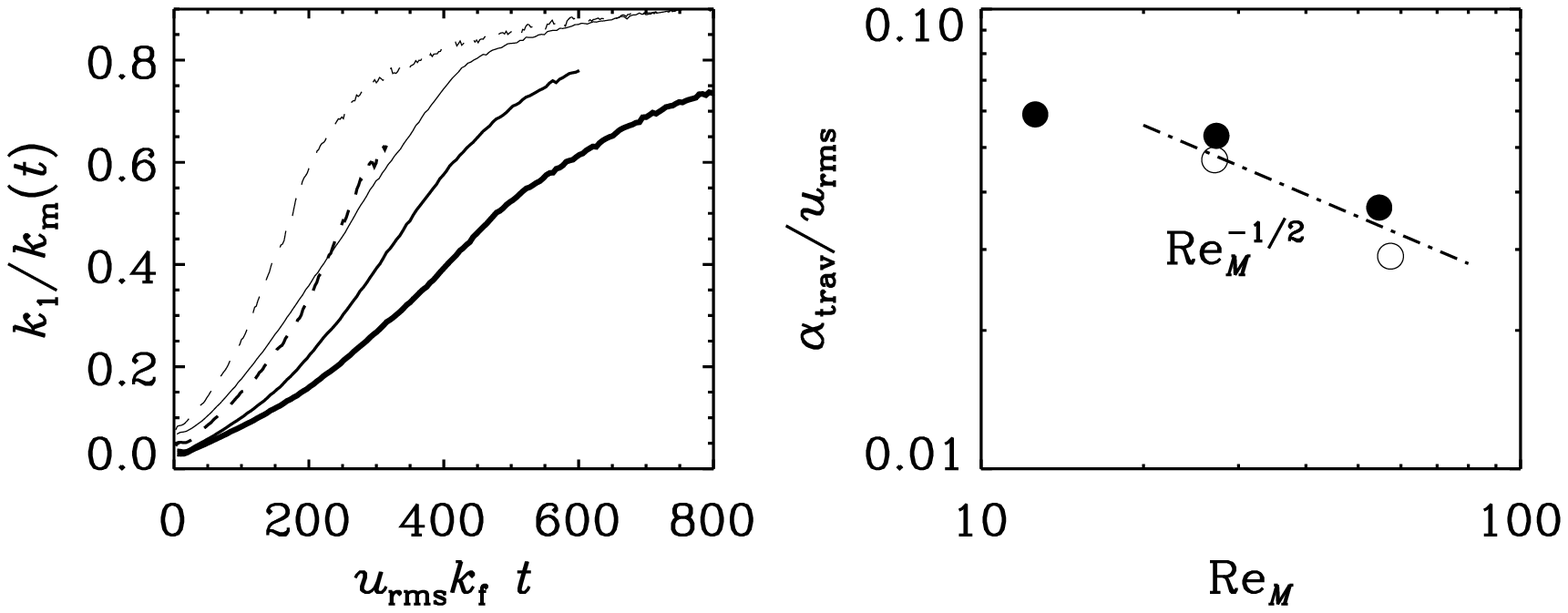}
\end{center}\caption[]{
Left panel: time dependence of the peak wavenumber for scale separation
ratios of 15 (dashed) and 30 (solid lines) at $\Rm$ of 12, 27, and
around 57 (increasing line thickness).
Right panel: $\Rm$ dependence of the cascade speed for scale separation
ratios of 15 (open symbols) and 30 filled symbols.
The straight lines give the $\Rm^{-1/2}$ (dotted) and
$\Rm^{-1}$ (dashed) dependences.
Adapted from \cite{B11_chandra}.
}\label{palp_trav}\end{figure}

\subsection{Inverse transfer and $\alpha$ effect}

The dynamics of a large-scale dynamo is most dramatic when the scale
separation ratio is large, i.e., $\kf/k_1\gg1$.
In \Fig{pspec_nohel512d2_pspec_PrM1} it was only 4, but now we consider
a case where $\kf/k_1 = 15$.
In \Fig{B} we show visualizations of one component of the field
for different times.
Evidently, a large-scale field develops that varies in the $y$ direction.
This particular large-scale field is best described by $xz$ averages.
In \Fig{psat} we show the evolution of the mean energy density of this
large-scale magnetic field, $\bra{\meanBB^2}$, and compare it with that of
the small-scale field, $\bra{\bb^2}=\bra{\BB^2}-\bra{\meanBB^2}$.
Note that the small-scale field reaches its final saturation value during
the time span considered, while the large-scale field has not yet saturated
and is expected to do so on the diffusive time scale of the box
such that $t\eta k_1^2=O(1)$.
It is also interesting to note that the large-scale field starts becoming
noticeable only when $t\eta k_1^2 \sim 0.02$, or
when $(t/t_{\rm d}) \sim 4.5$, where $t_{\rm d}=(\eta\kf^2)^{-1}$ is
the resistive timescale at the forcing scale.
In other words one needs several resistive (diffusive) times at the
forcing scale, before the large-scale field can grow. We will
interpret this result below in terms of the resistive
dissipation of small-scale magnetic helicity
which alleviates $\alpha$ quenching.

The evolution of the magnetic energy spectrum for this case is shown in
\Fig{pspec_ck}, where we see several stages during the early phase of
the dynamo, and in a separate panel the later saturation behavior.
Clearly, a large-scale field is already present for $t/\tau>100$,
but the field is then still fairly isotropic and therefore not
very pronounced in visualizations shown in \Fig{B}.
Spectra of magnetic energy and rescaled magnetic helicity are shown in
\Fig{pspec_ck0} for the saturated state.
Here, magnetic helicity spectra $\HM(k)$ are normalized such that
$\int\HM\,\dd k=\bra{\AAA\cdot\BB}$, where $\BB=\nab\times\AAA$
is the magnetic field expressed in terms of its vector potential.
Note that at early times, the magnetic field shows the Kazantsev $k^{3/2}$
slope ($t/\tau=10$) in the range $7\leq k/k_1\leq25$.
However, already at times $t/\tau=20$ and 30 one sees a small hump at
$\kf/2$, which is significant in view of an interpretation of these
results in terms of a so-called $\alpha^2$ dynamo, which will be discussed
in \Sec{MeanFieldTheory}.

The temporal  increase of the typical scale of the large-scale field
can be determined by monitoring the quantity
\EQ
\ell_{\rm m}(t)\equiv k_{\rm m}^{-1}(t)=
\left.\int k^{-1} \EM(k)\,\dd k\right/ \int \EM(k)\,\dd k.
\EN
In \Fig{palp_trav} we plot $k_{\rm m}^{-1}(t)$ for different values $\Rm$.
Here, we normalize with respect to the minimal wavenumber $k_1=2\pi/L$
in the domain of size $L$, so we plot $k_1/k_{\rm m}$.
There is a limited scaling range that allows us to determine the
increase of $\ell_{\rm m}(t)$ for different values of $\Rm$.
We measure the speed at with the bump travels toward larger scales
by the quantity $\alpha_{\rm trav}=\dd\ell_{\rm m}/\dd t$.
In the right-hand panel of \Fig{palp_trav} we plot the resulting
values of $\alpha_{\rm trav}$ as a function of $\Rm$.
The results are compatible with a resistively limited increase
whose speed diminishes like $\Rm^{-1/2}$.
This behavior was first seen in simulations of \cite{BDS02} and then
more convincingly at larger resolution in simulations of \cite{B11_chandra}.
Such a behavior further reinforces our earlier remark that
the large-scale field can only grow on the slow resistive timescale
in periodic boxes.

\subsection{Connection with mean-field theory}
\label{MeanFieldTheory}.

There exists a close analogy between the inverse transfer described
above and mean-field dynamo theory in that both are able to predict
the occurrence of large-scale fields with similar properties and
excitation conditions.
In mean-field theory one splits
the velocity field $\UU$ and magnetic field $\BB$
into the sum of a mean, large-scale components
($\meanUU$ and $\meanBB$)
and a turbulent, stochastic components ($\uu$ and $\bb$); that is
$\UU = \meanUU +\uu$ and $\BB = \meanBB +\bb$.
One then solves the averaged induction equation,
\EQ
{\partial\meanBB\over\partial t}=\nab\times\left(
\meanUU\times\meanBB+\meanEMF-\eta\mu_0\meanJJ\right),
\EN
where $\meanEMF=\overline{\uu\times\bb}$ is the mean electromotive
force that we discussed already in connection with \Eq{meanemf}.
Under the assumption of isotropy and sufficient scale separation in
space and time, we have just $\meanEMF=\alpha\meanBB-\etat\mu_0\meanJJ$,
where $\alpha$ and $\etat$ are a pseudo-scalar and a scalar respectively.
For the case when there is no mean flow, a
stability analysis gives the dispersion relation for the growth
rate $\lambda$ as
\EQ
\lambda=\alpha k-(\etat+\eta)k^2,
\EN
and the eigenfunctions are force-free solutions with
$\etat\mu_0\meanJJ=\alpha\meanBB$, which are plane polarized waves,
just like in \Fig{B}, where the large-scale field can be approximated
by $\meanBB\propto(\sin ky,0,\cos ky)$, ignoring here an arbitrary
phase shift in the $y$ direction.
The dynamo is excited when $C_\alpha\equiv\alpha/(\etat+\eta)k>1$, where
$C_\alpha$ is the relevant dynamo number in this context.
The fasted growing mode occurs at wavenumber
$k=k_{\max} = \alpha/[2(\etat+\eta)]$.
Furthermore, using estimates for the high-conductivity limit,
\EQ
\alpha=\alpK=-\onethird\tau\bra{\oo\cdot\uu} \quad {\rm and} \quad
\etat=\onethird\tau\bra{\uu^2},
\label{alpeta}
\EN
we find that $C_\alpha=\epsf\kf/k_1$
\citep{BDS02}, where $\epsf\leq1$ is the fractional helicity,
and $\eta\ll\etat$ has been assumed.
We can now return to our discussion in connection with \Fig{pspec_ck},
where we notice that at early times the field growth occurs at wavenumber
$\kf/2$.
This is indeed the value expected from our simple estimate,
since for fully helical turbulence, we expect
$\bra{\oo\cdot\uu} = \kf\bra{\uu^2}$, and thus
$k_{\max} = \alpha/2\etat = \kf/2$.

It is of interest at this point to comment on the validity
of mean-field dynamo concepts. In condensed matter physics for example
mean-field theory is generally valid when applied to systems
where fluctuations are assumed small. In high $\Rm$ turbulent systems
on the other hand, the fluctuations grow more rapidly than the
mean-field, due to small scale dynamo action. Thus, even in the
kinematic stage when Lorentz forces are small, one needs a closure
theory to calculate the mean-field
turbulent coefficients like $\alpha$ and $\etat$.
Traditionally these coefficients have
been derived by still making a quasilinear approximation (strictly
valid for small fluctuating fields), which is also known as the first order
smoothing approximation (FOSA) \citep{Mof78,KR80,BS05}.
This is sometimes also referred to as the second order correlation
approximation.
Some improvements to this can be made by adopting closure approximations
like the minimal tau approximation whereby triple correlations involving
fluctuating fields are taken to provide a damping term proportional to
the quadratic correlations \citep{BF,RKR,BS05}.
There are also a few cases, like $\delta$ correlated flows \citep{Kaz68,ZRS83}
or renovating flows \citep{Ditt,GB92} which provide analytically solvable
models, where the form of $\alpha$ and $\etat$ given by \Eq{alpeta} is
recovered.

In this context, direct simulations as discussed above, which can
be interpreted in terms of mean-field concepts, lend some validity
to the theory. This applies also to the interpretation
of results in the nonlinear regime to be discussed below.
Moreover, when the $\alpha$ and $\etat$ have been measured
directly in simulations of isotropic turbulence, one gets results
remarkably close to the estimates of FOSA given in \Eq{alpeta} \citep{SBS08}.
This suggests that the strong magnetic field fluctuations
produced by small-scale dynamo action do not
contribute a systematic large-scale component to the
mean emf $\meanEMF=\overline{\uu\times\bb}$, correlated
with the mean field. They do make the mean-field coefficients
noisy. However, the fact that we can still use the mean-field concept
in understanding the results of direct simulations implies
that this noise does not have a crucial effect, perhaps after
the small-scale dynamo has nearly saturated.
The saturation of the dynamo will be discussed further in \Sec{quench}.

\subsection{Shear dynamos}

Remarkably, not all large-scale dynamos require an $\alpha$ effect.
In fact, large-scale dynamo action has been seen in simulations with
just shear and no helicity; see \cite{B05} for simulations using
a shear profile motivated by that of the Sun.
An obvious candidate at the time was the so-called shear--current
effect \citep{RK03,RK04}, which requires $\eta_{ij}\meanU_{i,j}>0$,
where $\eta_{ij}$ is the part of the magnetic diffusivity tensor
that multiplies $\meanJJ$ such that $\meanemf_i=...-\eta_{ij}\meanJ_j$,
and $\meanU_{i,j}$ is the velocity shear.
Already early calculations using the test-field method showed that
the relevant component of $\eta_{ij}$ has the wrong sign \citep{B05b}.
This confirmed the results of quasilinear calculations
\citep{RS06,RK06,SSingh10,SSingh11}. Moreover in the large $\Rm$ limit
using FOSA, but for arbitrarily strong shear, the corresponding
cross coupling implied by the shear current effect was shown to be
absent \citep{SS09a,SS09b}. The issue of how the mean-field
grows in nonhelical turbulence in the presence of shear 
remained open in view of other possible contenders.

One possibility is the incoherent $\alpha$--shear dynamo that lives
from the combination of shear and fluctuations of the $\alpha$ effect
and was originally invoked by \cite{VB97} to explain the magnetic field
of unstratified shearing box simulations of accretion disc turbulence
\citep{HGB95}.
This mechanism has received considerable attention in the following
years \citep{Sok97,Sil00,Fed06,Pro07,KR08,SS09}.
Meanwhile, evidence for the existence of shear dynamos in simple
shearing box simulations was mounting \citep{Y08,Y08b,BRRK08,Rincon}.
Although the underlying mechanism may have appeared to be a new one,
there is now quantitative evidence that this can be explained by an
incoherent $\alpha$--shear dynamo \citep{BRRK08,Heinemann,MB12}.
This is remarkable given the unconventional nature of the approach
whereby one uses mean-field theory over two spatial directions and
considers the fluctuations that remain in time and the third coordinate
direction as physically meaningful.

\subsection{$\alpha$-quenching}
\label{quench}

A fully satisfactory theory for the magnetic feedback on the $\alpha$
effect does not exist.
What we do know is that for strong mean fields $\meanBB$,
this $\alpha$ becomes a tensor of the form
$\alpha_{ij}=\alpha_1(\meanB)\delta_{ij}
-\alpha_2(\meanB)\hatB_i\hatB_j$, where
$\hatBB=\meanBB/\meanB$ are unit vectors, and $\meanB=|\meanBB|$ is the modulus.
However, if this tensor is applied to the $\meanBB$ field, we have
$\alpha_{ij}\meanB_j=(\alpha_1-\alpha_2)\meanB_i$, which suggests that
$\alpha$ is just like a scalar.
We also know that at least part of the quenching acts in such a way
that the total field (small-scale and large-scale) obeys the magnetic
helicity evolution equation.
This was derived some time ago in a certain approximation by \cite{PFL}
and was then applied to derive an equation for the quenching of $\alpha$
\citep{KR82,GD94,Klee00,FB02,BB02,Sub02,BS05}.

\begin{figure}[t!]\begin{center}
\includegraphics[width=\columnwidth]{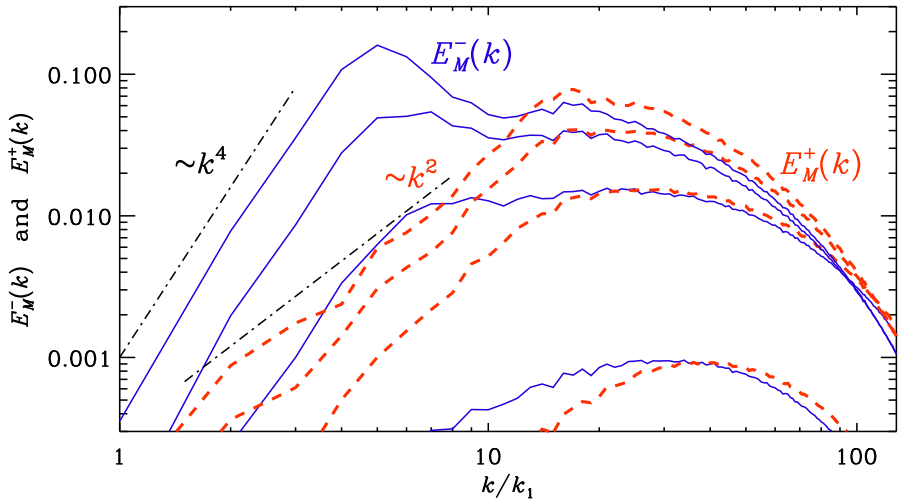}
\end{center}\caption[]{
Magnetic energy spectra $\EM^\pm(k)$ of positively (red)
and negatively (blue) polarized parts at earlier times.
Note the preferred build-up of the large-scale field at $\EM^-(\kf/2)$.
relative to the slower growth of $\EM^+(\kf)$.
Slopes proportional to $k^{2}$ for $\EM^+$ and
$k^{4}$ for $\EM^-$ are shown.
}\label{pspec_ck_pm}\end{figure}

The crucial starting point is the realization of \cite{PFL}
that under the influence of Lorentz forces,
the $\alpha$ effect has an additional component,
$\alpM=\onethird\tau\overline{\jj\cdot\bb}/\rho_0$, where
$\overline{\jj\cdot\bb}$ is the current helicity
associated with the small-scale field and $\alpha=\alpK+\alpM$
is the sum of kinetic and magnetic $\alpha$ effects.
Interestingly, \cite{PFL} also showed that $\etat$ does not get
renormalized under the same approximation.
Under locally isotropic conditions,
in the Coulomb gauge,
$\overline{\jj\cdot\bb}$ can be approximated by
$\kf^2\overline{\aaaa\cdot\bb}/\mu_0$, where
$\overline{\aaaa\cdot\bb}\equiv\meanhf$
is the magnetic helicity of the small-scale fields.
In order to write an evolution equation for the magnetic helicity
density one can fix a gauge for the vector potential.
One could also work in terms of the evolution
equation for the current helicity \citep{SB04}.
Perhaps more elegant is to write this evolution in terms of a gauge
invariant magnetic helicity density, defined as
the density of correlated links of $\bb$, and which
is most closely related to the $\meanhf$ in the Coulomb gauge \citep{SB06}.
The evolution equation for $\meanhf$ is
\EQ
{\partial\meanhf\over\partial t}=-2\meanEMF\cdot\meanBB
-2\eta\kf^2\meanhf-\nab\cdot\meanFFf,
\label{dhfdt}
\EN
where $\meanFFf$ is the magnetic helicity flux of the small-scale field.
This equation shows that the $\alpha$ effect produces magnetic helicity
at a rate $-2\meanEMF\cdot\meanBB=-2\alpha_{\rm red}\meanBB^2$, where
$\alpha_{\rm red}=\alpha-\etat\kmean$ is the reduced $\alpha$ effect and
$\kmean=\mu_0\meanJJ\cdot\meanBB/\meanBB^2$ is the effective wavenumber
of the mean field.
In a supercritical dynamo, the sign of $\alpha_{\rm red}$ agrees with
that of $\alpha$ (the $\etat$ term is subdominant).
Then starting with a specific sign for the
kinetic $\alpK$ and zero magnetic $\alpM$,
this produces $\alpM$ of opposite sign, which quenches 
the total $\alpha$ and the dynamo progressively with
increasing field strength.
In the absence of a magnetic helicity flux, this process happens on a
resistive time scale, which is what is seen in \Fig{psat}, where final
saturation is not even remotely in sight.
We recall that a rapid evolution of the energy of the mean field up
to $k_1/\kf$ times the equipartition value is expected on theoretical
grounds \citep{BB02}.
In practice, this is hard to verify because at early
times the mean field has not yet reached the scale of the system
and modes of different orientation are still competing.
Nevertheless, by splitting the magnetic field into its
positively and negatively polarized contributions,
$\EM^\pm(k)=\half[\EM(k)\pm k\HM(k)]$, it is possible to separate
large-scale and small-scale fields \citep{BDS02,BS05,B11_chandra}.
In \Fig{pspec_ck_pm}
we clearly see a faster build-up of the large-scale field through
$\EM^-(k)$ compared with the small-scale field through $\EM^+(k)$.
As we have argued before, this build-up of large-scale fields is still
resistively slow, but it is important to realize that the demonstrated
existence of large-scale fields in the kinematic stage provides support
for the usefulness of the mean-field approach.

The final saturation value, in periodic box simulations,
can be estimated simply by noting
that in the absence of magnetic helicity fluxes, the total
current helicity must vanish in the steady saturated case,
i.e., $\meanJJ\cdot\meanBB+\overline{\jj\cdot\bb}=0$.
This is because the total current helicity drives the
change of magnetic helicity, which in steady state must be zero.
Such a state can be obtained in a nontrivial manner with helical
forcing when current helicity has opposite signs at large and small
length scales.
For example, if the kinetic helicity at small scales is positive
(that is $\alpK$ is negative), then the generated $\alpM$ is positive.
In that case the current helicity of small-scale fields is also positive with
$\mu_0\overline{\jj\cdot\bb}=\kf\overline{\bb^2}$ and hence
$\mu_0\meanJJ\cdot\meanBB=-k_1\meanBB^2$ is negative.
(This implies that $\kmean=-k_1$.)
Assuming furthermore equipartition between kinetic and magnetic
energies at small scales, i.e.,
$\overline{\bb^2}\approx\mu_0\overline{\rho\uu^2}\equiv\Beq^2$,
we obtain \citep{B01}
\EQ
\meanBB^2/\Beq^2\approx\kf/k_1
\EN
in the final state.
We recall that in the run shown in \Fig{psat}, the scale separation
ratio is $\kf/k_1=15$, so it is understandable that there is not yet
any sign of saturation of the large-scale field; see \cite{B01} for
early results on the resistively slow saturation.

We do not expect resistively slow saturation behavior to occur in stars
or galaxies,
because the $\nab\cdot\meanFFf$ term can usually not be neglected
\citep{BF00a,BF00b,Klee00,BS05c,SSSB06}.
These results were obtained by solving the mean-field equations.
Subsequent simulations have shown that turbulent diffusive fluxes exist
that would constitute a sufficient contribution to $\nab\cdot\meanFFf$
\citep{Mitra10,HB10}, especially when $\Rm$ is larger
than a critical value around $10^4$.
This flux term can then dominate over the $2\eta\overline{\jj\cdot\bb}$
term \citep{Candel11}.
Under certain considerations it is possible that a flux of the form
$\meanEMF\times\meanAA$ from the mean electromotive force contributes
to the flux term, but by solving an evolution equation for the
magnetic helicity density of the total (small- and large-scale) field,
consideration of this term can be avoided.
This approach is also suited to deal with fluxes associated with
gauges that can introduce artificial fluxes in shearing environments;
see \cite{HB11}.
They find no evidence for a flux resulting from shear that were
previously argued to be important \citep{VC01}.
Using simulations with anisotropic non-helical
forcing in the presence of shear, \cite{SV11} argue that
large-scale dynamos might even live entirely due to helicity fluxes, 
although the exact origin of this flux remains to be clarified.
Another natural contribution to the flux term is just advection
of both small-scale and large-scale fields, along with the
associated magnetic helicity \citep{SSSB06,SB06}.
This will naturally arise from coronal mass ejections in the solar
context, or supernovae driven outflows in galaxies
\citep{BB03,SSSB06,SSS07,WBM11}.

Large-scale dynamos therefore seem to need helicity fluxes
to work efficiently.
This conclusion can be understood more physically as follows.
As the large-scale mean field grows, the turbulent emf $\meanEMF$
is transferring helicity between the small- and large-scale fields.
The large-scale helicity is in the links of the mean poloidal
and toroidal fields of the astrophysical system like the Sun or the Galaxy,
while the small-scale helicity is
in what can be described as `twist helicity' (or simply twist) of the
small-scale field, produced by helical motions.
Lorentz forces associated with the
`twisted' small-scale field would like to untwist the field.
This would lead to an effective magnetic $\alpM$ effect
which opposes the kinetic $\alpK$ produced by the
helical turbulence. The cancellation
of the total $\alpha$ effect can lead to
catastrophic quenching of the dynamo. This quenching can
be avoided if there is some way of dissipating the twist
(which is slow in high-$\Rm$ system) or transferring the twists
in the small-scale field out of the region of dynamo action,
or cancelling it between two hemispheres.
That is, if there are helicity fluxes out of the system or between
different parts of the system, the
large-scale field can grow to observable strengths.

\subsection{Application of mean-field theory to galaxies}

The mean-field theory described above has been applied extensively
to understand magnetic fields of disk galaxies.
The mean-field dynamo equations allow substantial simplification
provided a suitable parameterization of turbulent transport
coefficients is chosen. Of course, this parameterization presumes a
suitable closure for nonlinear effects to arrive at a closed set of
nonlinear mean-field dynamo equations. Such an approach does not necessarily
imply a deep understanding of the physical processes involved in the
magnetic field evolution.
However, it appears to be sufficient for pragmatic
modeling of magnetic field configurations in particular galaxies to
be compared with the observational data of polarized radio emission.

A first example of such parameterization and further drastic
simplification of mean-field dynamo equations was presented by
\cite{Par55}. Early simplified models for the galactic dynamo, which
allow analytic or quasi-analytic investigations, can be found in
\cite{RSS88}, and more recent reviews by \cite{Beck}, \cite{Shuk04},
and \cite{KZ08}.
Next, one can suggest, as the most pragmatic contender, simple mean-field
model for galactic dynamo the so-called no-$z$ model \citep{sm93,m95}.
The idea of this model is to present azimuthal and radial components
of the mean galactic magnetic field by their quantities at the galactic
equator and average the mean-field equations with respect to the
coordinate $z$ perpendicular to the galactic plane.
The third component of the magnetic
field can be reconstructed from the condition ${\rm div}
\meanBB=0$. This approach is an obvious oversimplification, but it
appears still adequate for modeling 
magnetic field evolution including the helicity fluxes discussed 
above \citep{SSS07,luke}. It also allows one to model
magnetic configurations for design
studies of new generations of radio telescopes such as the
Square Kilometer Array; cf.\ \cite{Metal12}.

\subsection{Application to mean-field dynamos of stars and the Sun}

In comparison with galactic dynamos, the status of the theory of solar
and stellar dynamos is less satisfactory.
Early models by \cite{SK69} provided numerical solutions to the full
two-dimensional axisymmetric mean-field equations under the assumption
of an assumed profile of the internal angular velocity $\Omega(r,\theta)$,
whose radial gradient, $\partial\Omega/\partial r$, was negative, and
variations in colatitude $\theta$ were ignored in most of their models.
This yielded cyclic magnetic fields with equatorward migration under
the assumption that the $\alpha$ effect is predominantly positive
in the northern hemisphere.
Subsequently, helioseismology delivered detailed contours of
$\Omega(r,\theta)$, which excluded the previously assumed $\Omega$
profiles and suggested that $\partial\Omega/\partial r>0$ at low
latitudes where strong magnetic flux belts are observed to propagate
equatorward during the course of the 11 year sunspot cycle.

Various solutions have been offered to this dilemma \citep{Par87}.
The most prominent one is the flux transport dynamo \citep{CSD95,DC99},
whereby meridional circulation causes the dynamo wave to turn around
in the opposite direction.
This model operates under the assumption that the turbulent magnetic
diffusivity operating on the toroidal field is much lower than the
value suggested by mixing length theory, and the $\alpha$ effect is
assumed to work only at the surface.
The other proposal is that the dynamo wave obeys equatorward
migration owing to a narrow near-surface shear layer \citep{B05},
where $\partial\Omega/\partial r$ is indeed strongly negative;
see Fig.~4 of \cite{bene}.
This proposal still lacks detailed modeling.
In view of the shortcomings in the treatment of mean-field dynamo
theory (e.g., our ignorance concerning nonlinearity discussed in
\Sec{quench} or the neglect of finite scale separation discussed below
in \Sec{ScaleSeparation}), the ground for speculation remains fertile.
Magnetic helicity fluxes in interface and flux transport dynamos
have already been looked at \citep{Chatterjee}, but finite scale
separation effects are neglected.
Indeed, the Sun is strongly stratified with its scale height changing
rapidly with depth, making it hard to imagine that simple-minded
approaches that ignore this can be meaningful at all.
A cornerstone for the proposal that solar activity is a shallow
phenomenon may lie in the success of explaining the formation of active
regions and perhaps even sunspots as a result of spontaneous formation
of flux concentrations by convective flux collapse \citep{KM00}
or the negative effective magnetic pressure instability; see
\Sec{stratification} below.

\subsection{Magnetic helicity in the solar dynamo exterior}

There is now explicit evidence for the presence of magnetic helicity
in the exterior of dynamos.
In particular, it has been possible to detect magnetic helicity of
opposite signs at small and large length scales.
This has been possible through measurements of magnetic helicity spectra
in the solar wind away from the equatorial plane \citep{BSBG11}.
This data pointed for the first time to a {\it reversal} of magnetic
helicity density between interior and exterior of the dynamo.
Such reversals have now been confirmed in numerical simulations of
dynamos coupled to an exterior.
There are first of all the simulations of \cite{WBM11} showing 
coronal mass ejections from a turbulent dynamo in a
wedge of a spherical shell, where the reversal occurred some distance
away from the dynamo.
Next, there are related simulations in Cartesian geometry where a reversal
can be found immediately above the surface; see figure~12 of \cite{WB10}.
Finally, there are earlier mean-field simulations showing such a reversal
as well; see the lower panel of figure~7 of \cite{BCC09}.

The occurrence of such a reversal is now well understood in terms
of the magnetic helicity equation for the small-scale field shown
in \Eq{dhfdt}.
In the dynamo interior, the $\alpha$ effect dominates over turbulent
magnetic diffusion and produces magnetic helicity of a sign opposite
to that of $\alpha$.
In the northern hemisphere, $\alpha$ is positive, so this produces
negative $\meanhf$.
Turbulent magnetic diffusivity opposes this effect, but it is still
subdominant and can therefore not change the sign of $\meanhf$.
This is different in the solar wind where the $\alpha$ becomes
subdominant compared with turbulent diffusion, which is the reason
the sign of $\meanhf$ is now different \citep{BSBG11}.

\subsection{Scale separation in space and time}
\label{ScaleSeparation}

In connection with \Eq{meanemf} we noted that the relation between
$\meanEMF$ and $\meanBB$ does, in general, involve a convolution in
space and time.
This becomes important when the variations of $\meanBB$ occur on time
and length scales comparable with those of the turbulence, whose turnover
time is $\ell/\urms$ and its typical scale is $\ell=\kf^{-1}$.
The properties of the integral kernel are often determined in Fourier
space, in which case a useful approximation of it is
$\hat\alpha_{ij}(k,\omega)=\alpha_{ij}^{(0)}\hat{K}(k,\omega)$ and
$\hat\eta_{ijk}(k,\omega)=\eta_{ijk}^{(0)}\hat{K}(k,\omega)$, where
\EQ
\hat{K}(k,\omega)={1\over1-\ii\omega\tau+k^2\ell^2}.
\label{Kkom}
\EN
Such an integral kernel has recently been obtained with the test-field
method applied to passive scalar diffusion \citep{RB12}, and in
limiting cases (either with $\omega=0$ or with $k=0$) for $\alpha$ and
$\etat$; see \cite{BRS08} and \cite{HB09} for details and applications
to spatial and temporal nonlocalities, respectively.
The test-field method allows one to determine the turbulent transport
coefficients by solving an extra set of equations that describe the
evolution of the fluctuating magnetic field for each test field,
which is a predetermined mean field.
Under some conditions it is also necessary to solve corresponding
evolution equations for velocity perturbations \citep{RB10}.
The combined presence of spatio-temporal nonlocality was first
considered by \citep{RB12}, who proposed \Eq{Kkom} and used it to
reformulate \Eq{meanemf} as a simple differential equation of the form
\EQ
\left(1+\tau{\partial\over\partial t}
-\ell^2{\partial^2\over\partial z^2}\right)\meanemf_i
=\alpha^{(0)}_{ij}\meanB_j+\eta^{(0)}_{ijk}\meanB_{j,k}.
\label{emf_nonlocal}
\EN
Such an equation is quite easy to implement and represents an
improvement in terms of physical realism.
This representation avoids not only the problem of causality
associated with the infinite speed of signal propagation in the
absence of the time derivative \citep{BKM04}, but it also prevents
the development of artificially small scales in the mean field.

The application of this new technique is still in its infancy,
and it needs to be seen to what extent spatio-temporal nonlocality
can substantially alter the nature of the solutions.
As an example we note that a finite $\tau$ has been found to lower
the critical dynamo number for oscillatory solutions by a factor of
about 2 \citep{RB12}.
In the context of disk galaxies, such non-locality in time
can also lead to phase shifts between the spiral forcing
of the dynamo by matter arms, and the resulting magnetic spirals,
as seen in the galaxy NGC6946 \citep{luke}.

\subsection{Magnetic structures resulting from strong density stratification}
\label{stratification}

Finally, let us discuss a mean-field effect that occurs
under the condition of strong density stratification.
It has been theoretically anticipated long ago
\citep{KMR93,KMR96,KR94,KRR89,KRR90,RK07},
but only more recently was it also seen in numerical simulations
of the mean-field equations \citep{BKR10} and then in direct
numerical simulations \citep{BKKMR11,KBKMR12}.

\begin{figure}[t!]\begin{center}
\includegraphics[width=\columnwidth]{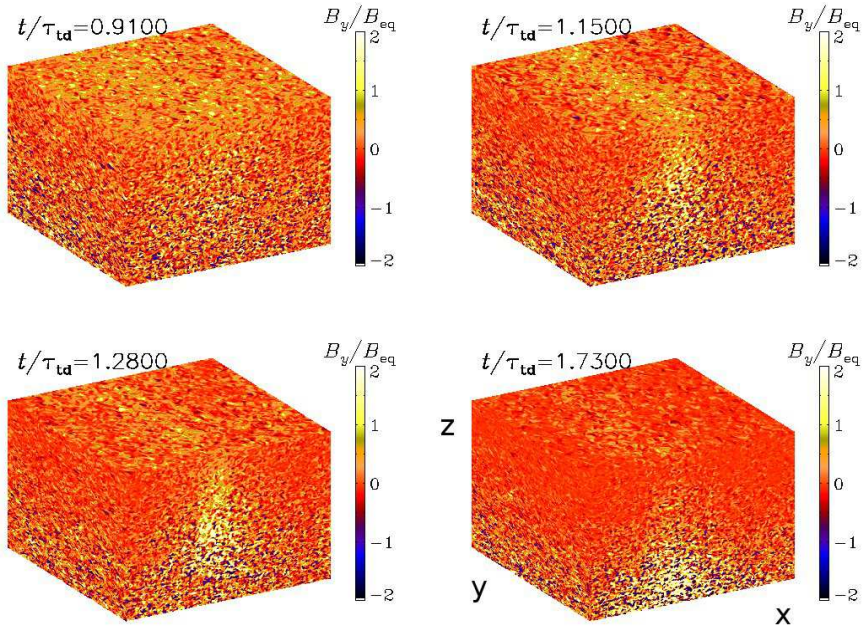}
\end{center}\caption[]{
Visualizations of $B_y$ for a simulation with strong density
stratification and a weak imposed magnetic field $(0,B_0,0)$
with $B_0/\Beqz=0.01$, $\Rm=18$ and $\Pm=0.5$.
Time is given in units of turbulent--diffusive times.
Note the gradual emergence of a large-scale magnetic flux concentration
which then sinks as a result of negative effective magnetic buoyancy.
Adapted from \cite{KBKMR12}.
}\label{img_1034}\end{figure}

The essence of this effect is the suppression of turbulent intensity
by a mean magnetic field.
This means that the effective pressure caused by $\meanB$ is not just
$\meanBB^2/2\mu_0$, but there must also be an additional contribution
from the suppression of the turbulence, which leads to a negative
contribution \citep{KRR89,KRR90,KR94,KMR96,RK07}.
This modifies the nature of the magnetic buoyancy instability in such
a way that magnetic structures become heavier than their surroundings
and sink.
This has been demonstrated using both mean-field simulations \citep{BKR10}
as well as direct numerical simulations \citep{BKKMR11}.
As an example we show in \Fig{img_1034} a snapshot from a direct
simulation where a weak ($B_0/\Beqz=0.01$) magnetic field is imposed
in the $y$ direction.
The density stratification is isothermal, so the density scale height
is constant in the direction of gravity (the negative $z$ direction).
The total density contrast from bottom to top is about 540.
The instability grows at a rate which scales with $\etatz k^2$,
where $\etatz=\urms/3\kf$ is an estimate for the turbulent magnetic
diffusivity, which is well reproduced by simulations using the
test-field method \citep{SBS08}.

The study of this negative effective magnetic pressure is still
very much in progress.
In particular, it has not yet been studied how this negative effective
magnetic pressure instability interacts with the mean-field dynamo.
It is envisaged that this instability might produce local magnetic
field enhancements in the surface layers that resemble active regions.
Such regions are long lived ($\sim1/2$ year).
Traditionally such long timescales have not been associated with
the surface layers.
However, the time scale of the negative effective magnetic pressure
instability is the turbulent--diffusive time, $\tautd=(\etatz k_1^2)^{-1}$,
where $\etatz=\urms/3\kf$ is the estimated turbulent magnetic diffusivity.
This time can be much longer than the local turnover time,
$\tauto=(\urms\kf)^{-1}$, which is about (1 day at the bottom of the
near-surface shear layer at 40 Mm depth.
The ratio of these time scales is $\tautd/\tauto=3(\kf/k_1)^2$, which can
be around 300 for a scale separation ratio of just 10.

In the rest of this paper we focus on small-scale magnetic
fields that occur over a range of different astrophysical settings.
They are believed to be important in understanding small-scale magnetic
fields in the surface layers of the Sun, although that part of the field
might also be a consequence of tangling the large-scale magnetic field.
Another possible astrophysical application of small-scale dynamos might
be the clusters of galaxies, because in the absence of rotation it is
difficult to motivate any form of large-scale dynamo action.

\section{Analytical approaches to small-scale turbulent dynamos}

Direct numerical simulations are now the most straightforward way to
understand turbulent dynamos. However, more traditional analytical
method provide a useful support for them.
Analytic methods have provided particular insights both into large-scale
and small-scale dynamos.
We discuss some specific 
analytic considerations of small-scale dynamos further below.

\subsection{Correlation tensor and small-scale dynamo}
\label{sec:1}

A natural approach here is to introduce the second-order correlation
tensor of the magnetic field $B_i ({\xx}, t)$ as
\begin{equation}
{\cal B}_{ij} ({\xx}, {\yy}, t_1, t_2) =
\bra{ B_i ({\xx}, t_1) B_j({\yy},t_2) },
\label{corr}
\end{equation}
taken at two spatial points ${\xx}$ and ${\yy}$ and at two instants
$t_1$ and $t_2$. Here $\bra{...}$ denotes averaging over an ensemble
of turbulent velocity field fluctuations which can be described by
the velocity field correlation tensor 
${\cal V}_{ij}({\xx}, {\yy}, t_1, t_2)$
constructed in the same way as ${\cal B}_{ij}$.

For a particular model of turbulence (short-correlated random flow),
\cite{Kaz68}, and simultaneously \cite{Kraichnan}
for a slightly different model, obtained a governing equation known
now as the Kazantsev equation.
In particular \cite{Kaz68} assumed
${\cal V}_{ij}({\xx}, {\yy}, t_1, t_2)= \bar{{\cal V}}_{ij}(\xx, \yy)
\delta(t_1-t_2)$, and derived an evolution equation for 
the magnetic field correlation tensor calculated at two simultaneous instants
${\cal B}_{ij}({\xx}, {\yy}, t) ={\cal B}_{ij}({\xx}, {\yy}, t, t)$, with
$t_1=t_2=t$. This reads
\begin{equation}
{{\partial {\cal B}_{ij}} \over {\partial t}} = \hat L_{ijkm} {\cal B}_{km}
\label{kas}
\end{equation}
where $\hat L_{ijkm}$ is a second-order differential operator with
coefficients depending on
$\bar{\cal V}_{ij}$, its spatial derivatives and
coefficient of magnetic diffusion $\eta$. In some sense,
the Kazantsev equation is similar to the famous
Steenbeck-Krause-R\"adler $\alpha$ effect equation in mean-field
electrodynamics. In practice however the latter equation provided
much more astronomically fruitful results than the first one. The
reason is presumably two-fold. First of all, the Kazantsev equation
requires more algebra for its
solution than the mean-field equations. We only take here some new
points isolated recently in this bulky algebra and refer to a
detailed review given by \cite{ZRS90}. The other reason is that the
physical interpretation of solutions of Eq.~(\ref{kas}) is
more delicate than that for the mean-field equation.
This is the main issue presented in following sections.

Original insight which lead Kazantsev to Eq.~(\ref{kas}) required
some mathematics from quantum field theory that can hardly be
considered `user friendly' for a person with an ordinary MHD background.
A more familiar approach, which is however rather bulky, can be
found elsewhere \citep{ZRS90,subamb,BS05}.
The particular form of the Kazantsev
equation (\ref{kas}) is associated with a specific model of
turbulence assumed, in particular the assumption of
$\delta$-correlated (in time) velocity fluctuations. This model is
quite restrictive and does not allow one to represent adequately
some basic properties of the Kolmogorov cascade. Various attempts to derive
this equations for more realistic models were undertaken
\citep[see, e.g.,][]{Kleeorin02}. For example, incorporating a finite
correlation time results in quite complicated albeit
beautiful mathematics and gives rise to integral equations. As far
as it is known however the results are more or less the same if just
applied to Eq.~(\ref{kas}) with $\bar{\cal V}_{ij}$ taken from a suitable
model of turbulence ignoring the fact that this very model is
incompatible with the derivation of Eq.~(\ref{kas}).

A reasonable way to simplify Eq.~(\ref{kas}) to a useful level is to
consider statistically homogeneous, isotropic and
mirror-symmetric turbulence and look for solutions having the same symmetry.
This means that we exploit a velocity correlation tensor of the form
\begin{equation}
\bar{\cal V}_{ij}= A(r)r_i r_j +  B(r) \delta_{ij}
\label{hom}
\end{equation}
(${\rr} = {\yy} - {\xx}$) and look for the magnetic
field correlation tensor in a similar form. The incompressibility
condition means that the functions $A$ and $B$ depend on a single
function, say, $F(r)$ while the solenoidality condition of the magnetic
field means that ${\cal B}_{ij}$ depends on a single function, say,
$W(r,t)$. Then Eq.~(\ref{kas}) can be reduced to a single second-order
ordinary differential equation for a single function which depends
on $W$ with coefficients depending on $F$. In fact it is the
equation which was obtained by \cite{Kaz68}. The algebra here
remains quite bulky and we avoid to present it here in detail; see,
for example, the detailed discussions in \cite{ZRS90} and \cite{BS05}.

There is no problem to solve the Kazantsev equation in the homogeneous
and isotropic case for a particular choice of $F$ numerically
or by analytical approximations. In fact the Kazantsev equation
can be reformulated as a Schr\"odinger type equation for a particle
with variable mass $m(r)$ in a potential $U(r)$ (both of which depend
on $F(r)$), whose bound states 
correspond to exponentially growing $W(r,t)$.
It is possible to develop a WKB-like method for an approximate
solution in the limit of large magnetic Reynolds numbers $\Rm$
($\sim \eta^{-1}$).

A general result following from these solutions can be summarized as
follows. For a sufficiently large $\Rm$, the magnetic field
correlation tensor (i.e.\ $W$) grows exponentially with a growth rate
$\gamma_2$ which is determined by $l/v$ where $l$ is the turbulence
correlation time and $v$ is its rms velocity. The critical magnetic
Reynolds number is of order $\Rmc \approx 10^2$, with
$\Rmc = 26$ in the most simple example with $F=\exp -(r/l)^2$.
For this illustrative example, function $W$ which determines
the magnetic field correlation properties has quite a complicated
form which contains spatial scales from $l$ up to $l\Rm^{-1/2}$.

A plausible scenario for the nonlinear saturation of the growth of the
magnetic field correlation tensor governing the Kazantsev equation
was suggested by \cite{S99}. The main idea here is that
nonlinearity results in an effective suppression of $\Rm$ up to
$\Rmc$. It is quite straightforward to consider
Kazantsev equation for homogeneous, isotropic and
mirror-asymmetric turbulence and to combine concepts of the second-order
correlation tensor with the mean-field approach \citep{S99,GS04}.
Interestingly, the Kazantsev equation 
in the presence of kinetic helicity, can be reformulated
into a tunnelling type quantum mechanical problem,
where by the bound states of the small scale dynamo can `tunnel' to 
develop long range correlations \citep{S99,BS00,BCR05}.
It is also possible to solve the Kazantsev equation
for locally homogeneous and isotropic turbulence in a finite body
of size $L \gg l$ \citep{Maslova,Belyanin}.

\subsection{Random magnetic fields in cosmology}

The description of a random magnetic field by a second-order correlation
function is widely exploited in cosmology. To be specific, \cite{Chernin}
considered at first a cosmological model with a random
magnetic field and discussed cosmological evolution of its magnetic
energy. We have to stress however that the concepts of statistical
hydrodynamics, which are a mathematical basis for the Kazantsev approach,
need some modification to be applicable for curved spaces of General
Relativity. The point is that Eq.~(\ref{corr}) considers the product
of two vectors applied at different spatial points while such option
is absent in Riemannian geometry. We have to consider a geodesic
line connecting points ${\xx}$ and ${\yy}$, transport a field
from the second point to the first one and consider product of
the two fields applied at one spatial point. This recipe presumes that a
geodesic lines which leaves a point $\xx$ does no longer cross
other geodesic lines which leave this point. In other words this
means that geodesic lines do not contain conjugated points and the
space-time has no gravitational lenses. If they do exist we have to
elaborate somehow the concept of the correlation tensor and no general
recommendations are suggested until now.

Then statistical homogeneity, isotropy and mirror symmetry in a
curved spatial section of a Friedmann cosmological model reads
\begin{equation}
{\cal B}^{ij}= C(r)n^i n^j + D(r) g^{ij} ,
\label{corr2}
\end{equation}
i.e.\ we have to distinguish upper and lower indices and use tangent
vector $n^i$ of the geodesic line connecting $\xx$ and $\yy$
instead of the vector $\rr$ which do not exist in a curved space.
The other point is that one
deals with a curved space formulation of
solenoidality condition to reduce $C$ and $D$ with one one function.
Inspired by earlier work of \cite{Andrade}, 
\cite{Rubashny} performed a corresponding analysis to
show that for the Lobachevsky space with negative curvature,
\begin{equation}
C= - R{\frac{{\rm th}\, {\frac {r}{R}}}{2}}F'\,, \quad \quad D=F+
R{\frac{{\rm th}\, {\frac {r}{R}}}{2}}F'\,.
\label{Lob}
\end{equation}
where $R$ is the curvature radius of the spatial section and $r$ is
distance between $\xx$ and $\yy$. It is instructive to compare
this representation with that for Euclidean space
\begin{equation}
C= - \frac{r}{2}F'\,, \quad \quad D=F+ \frac{r}{2}F'\,.
\label{Euc}
\end{equation}
Because ${\rm th}\, {\frac {r}{R}}$ has a finite limit at $r/R \to
\infty$, correlations decay slightly slower for the Lobachevsky
space rather in the Euclidean space (of course $F$ is the same in
both cases).
More significantly, however, the volume of a
sphere with radius $r$ grows exponentially in the Lobachevsky space
and not as a power law like in Euclidean space.
It means that $F$ has to decay much
faster in Lobachevsky space than in Euclidean space to get
convergence of various spatial means which are based on the
correlation tensor.

For a spherical space (closed cosmological model) one obtains
\begin{equation}
C= - R{\frac{{\rm tg}\, {\frac {r}{R}}}{2}}F'\,, \quad \quad D=F+
R{\frac{{\rm tg}\, {\frac {r}{R}}}{2}}F'\,.
\label{spher}
\end{equation}
This representation gives $C(\pi R) =0$, i.e., $C$ vanishes if
$\yy$ is just the opposite point to $\xx$. This looks reasonable
because $n^i$ is not determined uniquely for the opposite points.
${\rm tg}\, {\frac {r}{R}}$ diverges for $r = \pi R/2$ so the
finiteness of correlations means that $F'(\frac{\pi R}{2})=0$. This
condition is specific for spherical geometry and has no direct
analogues for the Euclidean one. Remarkably, a closed universe admits
another homogeneous and isotropic topological structure of spatial
section, namely an elliptical space which is twice smaller than the
spherical one. Instead of the above condition we find here $F''(\pi R/2)=0$.
Moreover, an elliptical space does not admit orientation, so one cannot
distinguish between left- and right-hand coordinate system there.
It means that quantities such as helicities, $\alpha$ effect
and other pseudo-scalar quantities cannot be introduced there. Such
topological constrains on the magnetic field correlation properties look
rather strange. Fortunately they do not affect substantially
physically interesting conclusions  because the correlation length
$l$ is usually much smaller than the curvature radius $R$.
The problem however is that both quantities as well as the horizon
radius vary strongly during the course of cosmological evolution that is
given $l$ being negligible in the present day cosmological scale
might be very large in scales of the Early Universe.
A more severe problem arises if we are going to consider a homogeneous and
isotropic ensemble of random gravitational waves \citep{Ivanova},
which are often discussed in cosmological context.

\subsection{Higher statistical moments and intermittency}

To get more detailed information concerning a dynamo-generated
small-scale magnetic field, it is useful to consider higher
statistical moments which are introduced as {\it ensemble} averages
of a product of $p$ magnetic field vectors ($p$ is the order of
statistical moment). Following the Kazantsev approach one can obtain the
governing equations for these quantities and demonstrate that the
moments grow
provided $\Rm$ is
high enough \citep[see, e.g.,][]{Kleeorin02}. Of course, the algebra
becomes more bulky as $m$ increases.
The problem is that the higher moments grow faster than the lower
ones in the sense that
\begin{equation}
\gamma_2/2 < \gamma_4/4 < \gamma_6/6 \dots
\label{mom}
\end{equation}
Of course, this fact can be supported by a direct calculation.
However, it is much more instructive to
demonstrate the phenomenon at a qualitative level
\citep[e.g.][]{Molchanov87}. Let us consider a flow with a memory
time $\tau$ so the magnetic field ${\BB} (n \tau)$ at instant $n \tau$
can be considered to be developed from the initial field
${\BB} (0)$ which is affected by $n$ independent random transport
operators $\hat T_i$
\begin{equation}
{\BB} (n \tau) = \hat T_n \hat T_{n-1} \dots T_1 {\BB}(0).
 \label{prod}
\end{equation}
As a matter of fact, progressive growth of higher statistical
moments in Eq.~(\ref{mom}) does not depend critically on the fine
structure of operators $\hat T_i$ so we can illustrate the
phenomenon by considering the simplest operators $\hat T_i$, i.e.\ just
independent random numbers $T_i$.

For the sake of definiteness, let $\ln T_i$ have Gaussian
distribution with zero mean and standard deviation $\sigma$. Then

\begin{equation}
T = T_n T_{n-1} \dots T_1
\label{mult}
\end{equation}
is a log-normal random quantity and $\ln T$ has zero mean and
standard deviation $\sqrt n \sigma = \sigma \sqrt{t/\tau}$. A
straightforward calculation shows that
\begin{equation}
\bra{T^p} = \exp (\sigma^2 p^2 t/2 \tau),
\label{aver}
\end{equation}
so the normalized growth rates of the moments $\gamma_p/p = \sigma^2
p/2 \tau$ grow linearly with the degree $p$ of statistical moment.
The other point is that the growth is determined not by a typical
value of $T_i$ which is of order $\sigma$, but by strong
deviations which are of order $\sigma \sqrt {tp/\tau}$. The probability
density of a Gaussian quantity to achieve the level $\sigma \sqrt
{tp/\tau}$ is of order $\exp (- tp/2 \tau)$ so the size $N$ of
a statistical ensemble should be as large as $N^*(t) = \exp (tp/2
\tau)$ to include such rare events.

Note that the above analysis presumes that the statistical ensemble
is infinitely large. If the ensemble is only large but finite its
size $N$ should exceed a critical value $N^*(t)$ which grows in time
exponentially. For large $t$ we obtain $N< N^* (t)$ and the above
estimates becomes inapplicable.

If we consider a medium of independent cells of size $l$
renovating after a memory time $\tau$, then the critical size of
the system which allows one to recognize the growth governed by
Eq.~(\ref{aver}) is given by
\begin{equation}
L^*(t) = l N^{1/3} = l \exp ( tp /6 \tau).
 \label{scale}
 \end{equation}
This means that the behavior of the statistical moment of the order
$p$ is determined by very rare cells and Eq.~(\ref{scale}) gives an
estimate for the distance to the nearest cell which determines the
moment at a given point. The phenomenon of a random field whose
properties are determined by rare and remote events is known as
intermittency. The wording comes from medicine and means a
state when a person is near death, but his/her heart still works
from time to time. These rare events of the heart activity determine
the fact that the person is still alive.

Note that if we calculated a PDF of $T$ based on a limited sample
with $N<N^*(t)$ it is practically impossible to recognize the
existence of the above mentioned rare events which do not contribute
to the PDF calculated. Of course, the importance of the result
depends on how large $t$ should be to make the intermittency
recognizable and how large the corresponding value $N^*$ is in
comparison with $N$ typical of celestial bodies.

It is natural to address this point based on a simple physical
example rather than just a product of random operators. A simple
example of this kind accessible for simple numerics has been
suggested by \cite{Zeldovich64} in a cosmological context. Let a
remote object have a (small) angular size $\theta$, let $x$ be the
distance to the object and $y =\theta x$ its linear size. $y (x)$
is known in the Riemannian geometry as Jacobi field and is governed
by the so-called Jacobi equation
\begin{equation}
y''+K y=0,
\label{jacobi}
\end{equation}
where a prime means the derivative taken with respect to $x$ and $K$ is the
spatial (sectional) curvature. \cite{Zeldovich64} recognized the
importance of density and then curvature fluctuations on the
evolution of $y$ along the line of sight.
In other words, we consider $K$ as a random, say, Gaussian quantity.

It is quite easy to simulate many independent
solutions of Eq.~(\ref{jacobi}) and determine experimentally how
large $N^*(t)$ is. It appears \citep{AS05} that one needs $N \approx
5 \times 10^5 \dots 10^6$ to recognize the difference between
$\gamma_2/2$ and $\gamma_4/4$ for $t\approx 10^2$. Of course, a
simulation of $10^6$ independent 3D cells for a hundred turnover times
becomes prohibitive given the purely computational problems, in addition
to the problems associated with the further data processing of the results. 
On the other hand, the number of independent turbulent
cells in a typical galaxy is of order of $10^6$ so the effects of
intermittency can contribute in mean quantities of interest for
galactic dynamos. In practice however the difference between
$\gamma_2/2$ and $\gamma_4/4$ is for the intermittent fields, as far
as it is known, not very large.

The other point here is that the growing magnetic field becomes
sooner or later dynamically important. Of course, it is important to know
which happens first, whether the field becomes dynamically important first
or whether the size of the domain becomes too small to contain
the intermittent structure?
Unfortunately, the abilities of simple models like
Eq.~(\ref{jacobi}) to reproduce the stage of nonlinear dynamo
saturation are limited. A numerical experiment with a simple model
shows that the number $N$ required to reproduce the behavior of
higher statistical moments declines strongly when the solution
becomes dynamically important such that exponential growth of
statistical moments saturates.

Yet another point which can be recognized from the experiences with
simple models is that ensemble averaging is not the only option to
describe the behavior of the growing solution.
\cite{Lamburt} demonstrated that the quantity
\begin{equation}
{{\ln |y|} \over t} \to \gamma >0 \quad \quad ({\rm for} \quad t \to
\infty)
\label{Furst}
\end{equation}
for almost all realizations (with probability 1, or ``almost sure'' in the
wording of probability theory) where no averaging is taken at all.
Quantities such as $\gamma$ are known as self-averaging quantities
and $\gamma$ in particular is known as Lyapunov exponent.
The phenomenon becomes clearer if one introduces a 2D row-vector $(y,y')$ and
rewrites Eq.~(\ref{jacobi}) as a vectorial equation
\begin{equation}
z' = z\hat A,
\label{unim}
\end{equation}
where $\hat A$ is a random matrix process with vanishing trace
(i.e.\ ${\rm Tr} \, \hat A =0$).
Then the evolution of $z$ from an initial condition $z_0$ can be
represented as a product of independent random unimodular matrices
$B_i = \exp (\hat A_i \tau)$, ${\rm det} \, B_i =  1$, where
$\hat A_i$ is a realization at a given interval of renovation
of the random matrix process $\hat A$.

The product of independent random matrices is quite well investigated in
probability theory (so-called Furstenberg theory).
\cite{Zeldovich84} stressed the importance of this theory for
small-scale dynamos (here the elements of $\hat A$ are $\partial v_i
/ \partial x_j$). \cite{Molchanov84} argue that magnetic field
generated by a small-scale dynamo grows such that
\begin{equation}
{{\ln |{\BB }|({\xx}, t)} \over t} \to \gamma >0 \quad \quad
{\rm for} \quad t \to \infty,
\label{dt}
\end{equation}
where $\gamma$ is a positive constant. A numerical experiment for the
Jacobi equation supports this interpretation and shows that $t$
should be of the order of a hundred memory times to get this behavior
\citep{AS05}. The PDF of the dynamo-excited magnetic
field is investigated by \cite{Chertkov}.

Unfortunately, this approach of simple models cannot mimic the
$\alpha$ effect. The point is that one cannot produce a
pseudoscalar quantity $\alpha$ based on correlations of $A_{ij}$
and $A_{mn}$ and a Levi-Civita  tensor $e_{pqr}$.

Generally speaking, the analytical results discussed above show
that detailed direct numerical simulations of small-scale dynamos at
the kinematic stage can be nontrivial to interpret due to
the rapidly growing intermittency. Fortunately, the magnetic field
becomes dynamically important quite rapidly (at least at small scales),
so the dynamo becomes nonlinear and mathematical
distinctions between properties of various statistical moments
become less and less important. \cite{AS05} investigated several
simple models how catastrophic intermittency typical of the
kinematic stage gradually evaporate when nonlinear effects become
important. Presumably, something like that happens for the much more
complicated full small-scale dynamo equations.

\subsection{Small-scale magnetic field and shell models}

The Kazantsev model is developed for the kinematic or weakly
nonlinear stage of the dynamo and its ability to describe the
strongly nonlinear stage of a dynamo is obviously quite limited.
However, it remains useful to have a simple model of strongly
nonlinear dynamos in terms of ordinary (instead of partial)
differential equations. An option of this kind is given by so-called
shell models of MHD-turbulence.

The starting point of the shell model approach is to note that for
random fields and flows, one can hardly reproduce in a numerical
simulation the actual realization of the random field which are obtained
in a given celestial body. In practice we are interested in some
spectral properties of the field of interest. If we can get such
properties without solving the full set of equations, we would be
happy with such a result.

Shell models are designed to describe the cascade process over a
large range of scales (wavenumbers) by a chain of variables
$u_n(t)$, $b_n(t)$, each of them characterizing velocity or
magnetic field oscillations with wavenumbers $k$ in the range from
$k_n=k_0 \lambda^n$ to $k_{n+1}$ i.e., a shell of wavenumbers. The
parameter $\lambda$ characterizes the ratio of two adjacent scales
(the width of the shell) and usually $\lambda \le 2$. The model
includes a corresponding set of ordinary differential equations,
which should reproduce the basic properties of the equation of motion.
In particular, the model has to reproduce the type of nonlinearity of the
primitive equations and to retain the same integrals of motion in the
dissipationless limit. Let us note that shell models can possess
positively defined integrals of motion (energy, enstrophy in
two-dimensional turbulence, and the square of magnetic potential in
2D MHD-turbulence), as well as quadratic integrals with an arbitrary
sign (the integrals of this kind are usually called `helicities').
The signs of the helicities are defined by the balance between the
contributions of odd and even shells to corresponding quantity.

The shell models were suggested by Kolmogorov's school to describe
the spectral energy transfer \citep{Gledzer,Desnianskii}.
After numerous refinements they became an effective tool for
description of the spectral properties of the small-scale turbulence
\citep[see for review][]{Bohr}. The shell models for MHD turbulence were
introduced by \cite{Frick84}, \cite{Gloaguen}, \cite{BEO96}, and \cite{FS98}.
This approach reveals many intrinsic features of small-scale
dynamo action in fully developed turbulence of conducting fluids
\cite[for review see][]{Biferale}. In particular, the shell model
suggested by \cite{FS98} gives a fast growth of small-scale magnetic
fields (on the timescale $l/v$) and its saturation at the
equipartition level as well as non-Gaussian (similar to lognormal)
PDF for small-scale magnetic field in the saturated state. In some
cases shell models give a hint concerning dynamo action in the
parametric domain inaccessible to direct numerical simulations.
In particular, \cite{Stepanov06} and \cite{Frick06} argue, based on
simulations of MHD-shell models, that the critical magnetic
Reynolds number for small-scale dynamo action remains moderate $\Rmc
\approx 80$ in the case of low Prandtl numbers.
This is also confirmed by the simulations discussed in \Sec{Low}.
General speaking, shell models seem to provide
an effective way to investigate small-scale dynamos.

It may be possible to combine shell models as a tool to describe
small-scale variables in a dynamo with grid or spectral methods for
large-scale variables in mean-field equations. However only the
first steps in this direction have been made until now
\citep{Frick02,Frick06,Nigro}.

\subsection{Dynamical chaos and small-scale dynamo}

One more point of comparison between numerical and analytical
approaches to the small-scale dynamos is as follows. The Kazantsev
model and its related investigations consider turbulence as a truly
random field and apply concepts of probability theory in full
extent. In contrast, direct numerical simulations and shell models
consider turbulence as a chaotic behavior of solutions for
deterministic equations of motion. It is quite risky to insist {\it
a priori} that dynamical chaos reproduces all properties of random
flows required for analytical approaches. It is even less obvious that
interstellar turbulence driven by supernova explosions
\citep[e.g.][]{Korpi,Gressel}
provides a truly random velocity field. On the other hand, dynamo models based
on steady flows with stochastic flow lines such as ABC flows
excite magnetic fields which look rather different from the field
discussed for turbulent dynamos. Zeldovich \citep[see][]{ZRS83}
suggested that dynamo action in nonstationary flows (say, when
parameters $A$, $B$ and $C$ in the ABC flow fluctuate in time)
becomes much more similar to the dynamo action in random flows than
dynamos in stationary flows. Recent work of \cite{Kleeorin09}
supports this idea and demonstrates that in a fluctuating
ABC flow, a large-scale magnetic field can indeed grow in a way
similar to what is supposed to grow in a random
mirror-asymmetric (helical) turbulent flow.

\section{Conclusions}

In this review we have discussed our current understanding of both
large-scale and small-scale dynamos that are relevant in astrophysics.
In particular, we have illustrated differences and similarities
between them and have compared them in terms of the energy spectra
with the corresponding cases at low magnetic Prandtl numbers.
We have also briefly highlighted the resistively slow saturation
phenomenon as well as catastrophic quenching of helicity-driven
large-scale dynamos.
Finally, we have discussed new issues in connection with small-scale
dynamos and their intermittency.

Relative to earlier reviews \citep[e.g.][]{BS05} there have been some
unexpected advances regarding the nature of magnetic helicity fluxes
and direct observational evidence for magnetic helicity in the solar wind.
Another completely unexpected development concerns the numerical detection
of the negative effective magnetic pressure instability in simulations of
strongly stratified turbulence.
It is expected that these developments will contribute to an improved
understanding of the magnetic field generation in astrophysical bodies.
On the technical side, \cite{BS05} discussed just the basics of the
basics of the test-field method, but now this technique has provided
significant insights into issues such as non-locality in space and
time, as well as the nonlinear quenching of dynamo coefficients.

In this review we have barely touched upon applications to actual
astrophysical bodies.
In fact, a lot of progress has been made by trying to model the Sun.
Direct numerical simulations of convection in spherical
shells has shown signs of cyclic large-scale fields
\citep{BBBMT10,BMBBT11,KKBMT10,Ghizaru,Racine}, but only for
systems that are rotating at least 3 times faster than the Sun;
see \cite{BMBBT11} for simulations with otherwise realistic
solar parameters.
Similar results have also been obtained for local simulations
of the galactic dynamo, which only appears to be excited when
the rotation speed is artificially enhanced \citep{Gressel}.
This might well indicate that one is on the right track, but that
the turbulence present in the system is exerting too much effective
diffusion owing to it being dominated by rather large eddies.
It is conceivable that in reality the turbulent eddies would
be smaller, lowering thereby the effective turbulent diffusion,
which can at the moment (with the larger eddies) only be emulated
by adopting faster rotation.
Similar results have recently also been seen in models of
the negative effective magnetic pressure instability, where
direct numerical simulations showed the development of the
instability only when there were enough turbulent eddies in
the domain and thereby the turbulent diffusivity sufficiently
small on that scale; see \cite{BKKMR11} and, in particular,
Figure~17 of \cite{BKKR12}.
Thus, much has been learned about turbulent dynamos and their
relevance for astrophysical systems but as usual, much remains
to be elucidated.

\begin{acknowledgements}
We thank Andre Balogh for providing an inspiring atmosphere at the
International Space Science Institute in Bern in 2010, which has led
to new collaborations and scientific progress.
Computing resources were provided by the Swedish National Allocations Committee
at the Center for Parallel Computers at the Royal Institute of Technology in
Stockholm and the High Performance Computing Center North in Ume{\aa}.
This work was supported in part by the European Research Council
under the AstroDyn Research Project No.\ 227952
and the Swedish Research Council under the project grant 621-2011-5076.
\end{acknowledgements}


\end{document}